%% file: 2DFishnetsBD.tex
\documentclass[12pt,a4paper,sort&compress]{article}

\usepackage[T1]{fontenc}
\usepackage[latin1]{inputenc}
\usepackage[nosort]{cite}
\usepackage{color}
\usepackage{graphicx}
\usepackage{bbm}
\usepackage{amsmath}
\usepackage{amssymb}
\usepackage{physics}
\usepackage{mathtools}
\usepackage{subfig}
\usepackage{verbatim}
\usepackage{multirow}
\usepackage{tikz}
\usepackage{amsthm}
\usepackage{fancyhdr}
\usepackage{wrapfig}
\usepackage{hyperref}
\usepackage{dsfont}
\usepackage{delimset} 
\usepackage{shuffle} 
\usepackage{mathtools}
\usepackage{todonotes}
\input cyracc.def

\makeatletter \@addtoreset{equation}{section} \makeatother

\makeatletter
\let\old@startsection=\@startsection
\let\oldl@section=\l@section
\renewcommand{\@startsection}[6]{\old@startsection{#1}{#2}{#3}{#4}{#5}{#6\mathversion{bold}}}
\renewcommand{\l@section}[2]{\oldl@section{\mathversion{bold}#1}{#2}}
\makeatother

\makeatletter
\let\old@makecaption=\@makecaption
\def\@makecaption{\small\old@makecaption}
\makeatother

\let\oldPhi=\Phi
\let\oldPsi=\Psi
\let\oldGamma=\Gamma
\let\oldDelta=\Delta
\let\oldSigma=\Sigma
\let\oldTheta=\Theta
\let\oldPi=\Pi
\let\oldUpsilon=\Upsilon
\renewcommand{\Phi}{\mathnormal{\oldPhi}}
\renewcommand{\Psi}{\mathnormal{\oldPsi}}
\renewcommand{\Gamma}{\mathnormal{\oldGamma}}
\renewcommand{\Sigma}{\mathnormal{\oldSigma}}
\renewcommand{\Delta}{\mathnormal{\oldDelta}}
\renewcommand{\Theta}{\mathnormal{\oldTheta}}
\renewcommand{\Pi}{\mathnormal{\oldPi}}
\renewcommand{\Upsilon}{\mathnormal{\oldUpsilon}}

\newcommand{\superN}{\mathcal{N}}

\newcommand{\mySigma}{\sigma}

\makeatletter
\newlength{\apb@width}
\newcommand{\autoparbox}[2][c]{\settowidth{\apb@width}{#2}\parbox[#1]{\apb@width}{#2}}

\makeatother

\ifx\genfrac\sdflkaj

\else

\fi

\newcommand{\mV}{\mathcal{V}}

\newcommand{\oG}{{\overline{\mathcal{G}}}}
\newcommand{\cG}{{\mathcal{G}}}

\newcommand{\grp}[1]{\mathrm{#1}}

\makeatletter
\def\mr@ignsp#1 {\ifx\:#1\@empty\else #1\expandafter\mr@ignsp\fi}%
\newcommand{\multiref}[1]{\begingroup
\xdef\mr@no@sparg{\expandafter\mr@ignsp#1 \: }%
\def\mr@comma{}%
\@for\mr@refs:=\mr@no@sparg\do{\mr@comma\def\mr@comma{,}\ref{\mr@refs}}%
\endgroup}
\makeatother

\newcommand{\hypref}[2]{\ifx\href\asklfhas #2\else\href{#1}{#2}\fi}

\renewcommand{\eqref}[1]{(\multiref{#1})}

\ifx\href\asklfhas\newcommand{\href}[2]{#2}\fi

\DeclareMathOperator{\Sol}{Sol}

\DeclareMathOperator{\CPFI}{\mathsf{CYPFI}}

\newcommand{\ua}{\underline{a}}

\newcommand{\uPi}{\underline{\Pi}}

\newcommand{\ux}{\underline{x}}
\newcommand{\uw}{\underline{w}}
\newcommand{\uz}{\underline{z}}
\newcommand{\uq}{\underline{q}}

\newcommand{\rd}{\mathrm{d}}

\def\beq{\begin{equation}}
\def\eeq{\end{equation}}
\def\bsp#1\esp{\begin{split}#1\end{split}}

\usepackage{color}

\DeclareMathOperator{\K}{K}

\DeclareMathOperator{\Sym}{Sym}
\DeclareMathOperator{\End}{End}

\begin{document}

\thispagestyle{empty}

\begin{flushright}\footnotesize
\texttt{BONN-TH-2023-08}\\
\texttt{TUM-HEP-1470-23}\\
\end{flushright}
\vspace{.2cm}

\begin{center}%
{\LARGE\textbf{\mathversion{bold}%
The Basso-Dixon Formula and \\Calabi-Yau Geometry}\par}

\vspace{1cm}
{\textsc{Claude Duhr${}^a$, Albrecht Klemm${}^{a,b}$, Florian Loebbert${}^a$,\\ Christoph Nega${}^c$, Franziska Porkert${}^a$ }}
\vspace{8mm} \\
\textit{%
${}^a$Bethe Center for Theoretical Physics, Universit\"at Bonn, D-53115, Germany\\[2pt]
${}^b$Hausdorff Center for Mathematics, Universit\"at Bonn, D-53115, Germany\\[2pt]
${}^c$Physics Department, Technical University of Munich, D-85748 Garching, Germany
}
\vspace{.5cm}

\texttt{\{cduhr,aoklemm,loebbert,fporkert\}@uni-bonn.de}\\[2pt]
\texttt{c.nega@tum.de}
%

\par\vspace{15mm}

\textbf{Abstract} \vspace{5mm}

\begin{minipage}{13cm}

We analyse the family of Calabi-Yau varieties attached to four-point fishnet integrals in two dimensions. We find that the Picard-Fuchs operators for fishnet integrals are exterior powers of the Picard-Fuchs operators for ladder integrals. This implies that the periods of the Calabi-Yau varieties for fishnet integrals can be written as determinants of periods for ladder integrals. The representation theory of the geometric monodromy group plays an important role in this context. We then show how the determinant form of the periods immediately leads to the well-known Basso-Dixon formula for four-point fishnet integrals in two dimensions. Notably, the relation to Calabi-Yau geometry implies that the volume is also expressible via a determinant formula of Basso-Dixon type. Finally, we show how the fishnet integrals can be written in terms of iterated integrals naturally attached to the Calabi-Yau varieties.

\end{minipage}
\end{center}

\newpage
\tableofcontents
\bigskip
\hrule

\newpage


\input{introduction}

\input{setup}



\input{4-pt}

\input{volumes}

\section{Conclusions}
\label{sec:conclusions}

In this paper, we have studied the Calabi-Yau geometries $\mathcal{M}_{M,N}$ that arise in the context of the four-point fishnet integrals $I_{M,N}$ in $D=2$ dimensions, where the internal vertices are arranged on a $M\times N$ square lattice. We have focused, in particular, on the Picard-Fuchs operators of $\mathcal{M}_{M,N}$, which encode the information on the periods. We find that, remarkably, these Picard-Fuchs operators can be written as the $M^{\textrm{th}}$ exterior power of the ladder integral with $W=M+N-1$ loops. This implies that all periods of $\mathcal{M}_{M,N}$ can be written as determinants of periods of $\mathcal{M}_{1,W}$. We have studied in detail relations among these determinants. The exterior power structure also allows us to determine the intersection form on $\mathcal{M}_{M,N}$. The periods together with the intersection form provide enough information to compute $I_{M,N}$, and remarkably the solution naturally takes the form of a determinant. This provides a possible geometric origin of the Basso-Dixon formula for fishnet integrals in $D=2$ dimensions. We have also studied the quantum volume of the mirror Calabi-Yau, and we find that it can naturally be expressed in terms of iterated integrals.

For the future, it would be interesting to extend our work into two possible directions. First, currently our results only apply to four-point fishnet integrals where all vertices attached to the same side of the square formed by the internal vertices have been identified. Since our results show that the Basso-Dixon formula for these integrals is a consequence of the structure of the Picard-Fuchs operator, it would be interesting to understand if a similar structure is present in fishnet integrals depending on more external points. In this case, the periods are solutions to an ideal of partial differential equations, which complicates the analysis. A set of generators of this ideal is conjectured to be given by the generators of the Yangian over the conformal algebra~\cite{Duhr:2022pch}. 
Second, it would be interesting to understand if also the Basso-Dixon formula in $D=4$ dimensions has a similar mathematical origin. While in four dimensions there is no connection to Calabi-Yau geometry, this is not a necessary condition, and it may be sufficient to find a differential operator that can be written as an exterior power. An intermediate step towards these goals could be to extend the results of this paper to the four-point fishnet integrals with more general propagator powers considered in ref.~\cite{Derkachov:2018rot}. We leave these questions for future work.


\section*{Acknowledgements}
The authors thank V.~Kazakov. We want, in particular, to thank Matt Kerr for his visit to the MPIM in Bonn and the discussions on representation theoretic  questions in Hodge Theory. CN was supported by the Excellence Cluster ORIGINS funded by
the Deutsche Forschungsgemeinschaft (DFG, German Research Foundation) under Germany's Excellence Strategy - EXC-2094 - 390783311. 
The work of FL is supported by funds of the Klaus Tschira Foundation gGmbH.
This work was co-funded by
the European Union (ERC Consolidator Grant LoCoMotive 101043686 (CD, FP) and ERC Starting Grant 949279
HighPHun (CN)). Views and opinions expressed are
however those of the author(s) only and do not necessarily reflect those of the European
Union or the European Research Council. Neither the European Union nor the granting
authority can be held responsible for them.

\appendix
\input{app_proof_det}

\input{app_proof_rels}
\input{app_proof_IIQ}

\bibliographystyle{nb}
\bibliography{2DFishnetsBD}

\end{document}

%% file: introduction.tex

\section{Introduction}

Over the last decade, it has become clear that the analytic structure of perturbative scattering amplitudes and multi-loop Feynman integrals is tightly related to topics in algebraic geometry. In particular, it is known that Feynman integrals compute (relative) periods in the sense of Kontsevich and Zagier~\cite{MR1852188,Bogner:2007mn}. As a consequence, understanding the geometry associated to a Feynman integral may inform us about the class of transcendental functions and numbers that appear in the result, and methods for the computation of periods may be adapted to perturbative computations in quantum field theory. Which classes of geometries and functions may arise from Feynman integrals is still an open question. The simplest examples can either be expressed in terms of polylogarithmic functions or involve elliptic or modular curves. The corresponding functions are by now relatively well understood (see, e.g., ref.~\cite{Bourjaily:2022bwx} and references therein for a recent review). It is known that also higher-genus Riemann surfaces may show up~\cite{Huang:2013kh,Hauenstein:2014mda,Marzucca:2023gto}, though in that case the relevant functions are still poorly understood~\cite{DHoker:2023vax}. In addition, Calabi-Yau varieties may arise, and in this case the relevant class of functions is slowly emerging, see, e.g., refs.~\cite{Brown:2010bw,Bloch:2014qca,MR3780269,Bourjaily:2018ycu,Bourjaily:2018yfy,Bourjaily:2019hmc,Klemm:2019dbm,Bonisch:2020qmm,Bonisch:2021yfw,Duhr:2022pch,Duhr:2022dxb,Pogel:2022yat,Pogel:2022ken,Pogel:2022vat,McLeod:2023doa}. 

Particularly interesting representatives of Feynman integrals which involve Calabi-Yau geometries are the \emph{ladder}, \emph{traintrack} and \emph{fishnet integrals}~\cite{Bourjaily:2018ycu,Bourjaily:2018yfy,Bourjaily:2019hmc,Duhr:2022pch,McLeod:2023doa}, because they compute correlators in the so-called {fishnet conformal field theories} of refs.~\cite{Gurdogan:2015csr,Kazakov:2018qbr,Kazakov:2022dbd}. 
Originally, these fishnet theories were discovered in $D=4$ spacetime dimensions, where they arise as double-scaling limits of planar gamma-deformed $\superN=4$ Super Yang-Mills theory with gauge group $\grp{SU}(N_\text{c})$. The gamma-deformed theory depends on three parameters $\gamma_{1,2,3}$ in addition to the Yang-Mills coupling constant~$g$. In its simplest version, the fishnet limit of this model is defined by taking $g\to 0$ and $\gamma_3\to i\infty$, while $\xi^2:=g^2N_\text{c}e^{-i\gamma_3}$ is kept fixed and furnishes the new coupling constant of the limit theory. 
Similar to $\superN=4$ Super Yang-Mills theory, the fishnet theory has an 
AdS/CFT dual, the so-called fishchain model introduced in refs.~\cite{Gromov:2019aku,Gromov:2019bsj}. Different combinations of limits in the above parameters lead to more involved families of fishnet theories, which can in turn be generalised, e.g.\ to generic spacetime dimensions $D$.
Notably, in the planar limit the fishnet models inherit integrability properties from their AdS/CFT mother theories, which allows one to study Feynman integrals as integrable systems~\cite{Zamolodchikov:1980mb,
Basso:2017jwq,
Chicherin:2017frs,
Gromov:2017cja,
Basso:2018agi,
Derkachov:2018rot,
Gromov:2018hut,
Loebbert:2019vcj,
Derkachov:2019tzo,
Loebbert:2020hxk,
Gurdogan:2020ppd,
Basso:2021omx,
Cavaglia:2021mft,
Loebbert:2022nfu,
Kazakov:2023nyu,
Aprile:2023gnh,
Kade:2023xet}.

Fishnet integrals are also interesting purely from the Feynman integral perspective, because they have a particularly simple analytic structure, which often allows one to obtain analytic expressions even at high loop orders. In $D=4$ dimensions, it is known that all four-point ladder integrals can be evaluated in terms of classical polylogarithms~\cite{Usyukina:1992jd,Usyukina:1993ch}.  Analytic results are also known for two-loop traintrack integrals in terms of elliptic polylogarithms~\cite{Wilhelm:2022wow,Kristensson:2021ani,Morales:2022csr}. Remarkably, it was conjectured in ref.~\cite{Basso:2017jwq}, and later proven in refs.~\cite{Basso:2021omx,Aprile:2023gnh}, that four-point fishnet integrals where the internal vertices are arranged on a square lattice can be expressed as determinants of ladder integrals. This simple analytic structure, known as the \emph{Basso-Dixon formula}, was also observed for four-point fishnet integrals in $D=2$ dimensions in ref.~\cite{Derkachov:2018rot}. 

In ref.~\cite{Duhr:2022pch} we argued that fishnet integrals in two dimensions are tightly related to certain families of Calabi-Yau (CY) varieties, and that the value of the fishnet integral computes  the \emph{quantum volume} of the mirror CY geometry. One of the goals of this paper is to study the relevant Calabi-Yau geometries in the context of four-point fishnet integrals. By conformal invariance, these integrals depend on a single cross ratio, and correspondingly the family of Calabi-Yau varieties is described by a single complex modulus. A lot of information about a Calabi-Yau variety is encoded in its periods. Due to conformal invariance, in the case of four-point fishnet integrals we need to consider a one-parameter family of Calabi-Yau varieties, and the periods are given by solutions of an ordinary differential operator, called the Picard-Fuchs operator. We present a strategy to determine the Picard-Fuchs operator for fishnet integrals. We observe that for ladder integrals, the Picard-Fuchs operators fall into the class of so-called \emph{Calabi-Yau operators}~\cite{Almkvist1,Almkvist2,BognerCY,BognerThesis}. In the case of a non-ladder integral, the Picard-Fuchs operator is the exterior power of the differential operator for a ladder integral. 
This implies that the periods can be written as determinants, very reminiscent of the Basso-Dixon formula. Remarkably, this also shows that fishnet integrals and volumes can be written in a Basso-Dixon-like determinantal form. Since the solutions of the ladder Picard-Fuchs operators 
form irreducible representations of the geometric monodromy group, representation theory 
determines to a large extent the properties of the solutions for the general fishnet operators.   
For the $\ell$-loop ladder integrals we employ a Picard-Fuchs operator of order $\ell+1$, while the order of the differential operator associated to generic Basso-Dixon graphs is given by $b_\ell$ as specified in table~\ref{tab:h_p} in section~\ref{sec:periodsgeneral}.

This paper is organised as follows: In sections~\ref{sec:setup},~\ref{sec:CY_FN} and~\ref{sec:DEQ_ops} we give a short review of fishnet integrals,  Calabi-Yau geometry and general properties of Picard-Fuchs operators. In particular in section \ref{sec:defvolumes} we explain how the fishnet integrals are related to the \emph{calibrated volume} of the Calabi-Yau manifold ${\cal M}$ or the \emph{quantum volume} of its mirror ${\cal W}$. In section~\ref{sec:PF_fishnets} we analyse the Picard-Fuchs operators for the Calabi-Yau varieties obtained from fishnet integrals, and we discuss the properties of their periods. In section~\ref{sec:BD} we show how to compute the fishnet integrals from these periods, and we demonstrate how the Basso-Dixon formula arises naturally from our geometric considerations. In section \ref{sec:volumes}
we show how the \emph{volumes} of the Calabi-Yau varieties can naturally be expressed in terms of iterated integrals, which can themselves be cast in the form of a Basso-Dixon-like formula. Finally, in section~\ref{sec:conclusions} we present our conclusions. We include several appendices with mathematical proofs omitted throughout the main text.

%% file: setup.tex

\section{Fishnet integrals and Calabi-Yau varieties}
\label{sec:setup}

\subsection{Fishnet integrals and the Basso-Dixon formula}

\begin{figure}[t]
\centering
\includegraphics[scale=0.25]{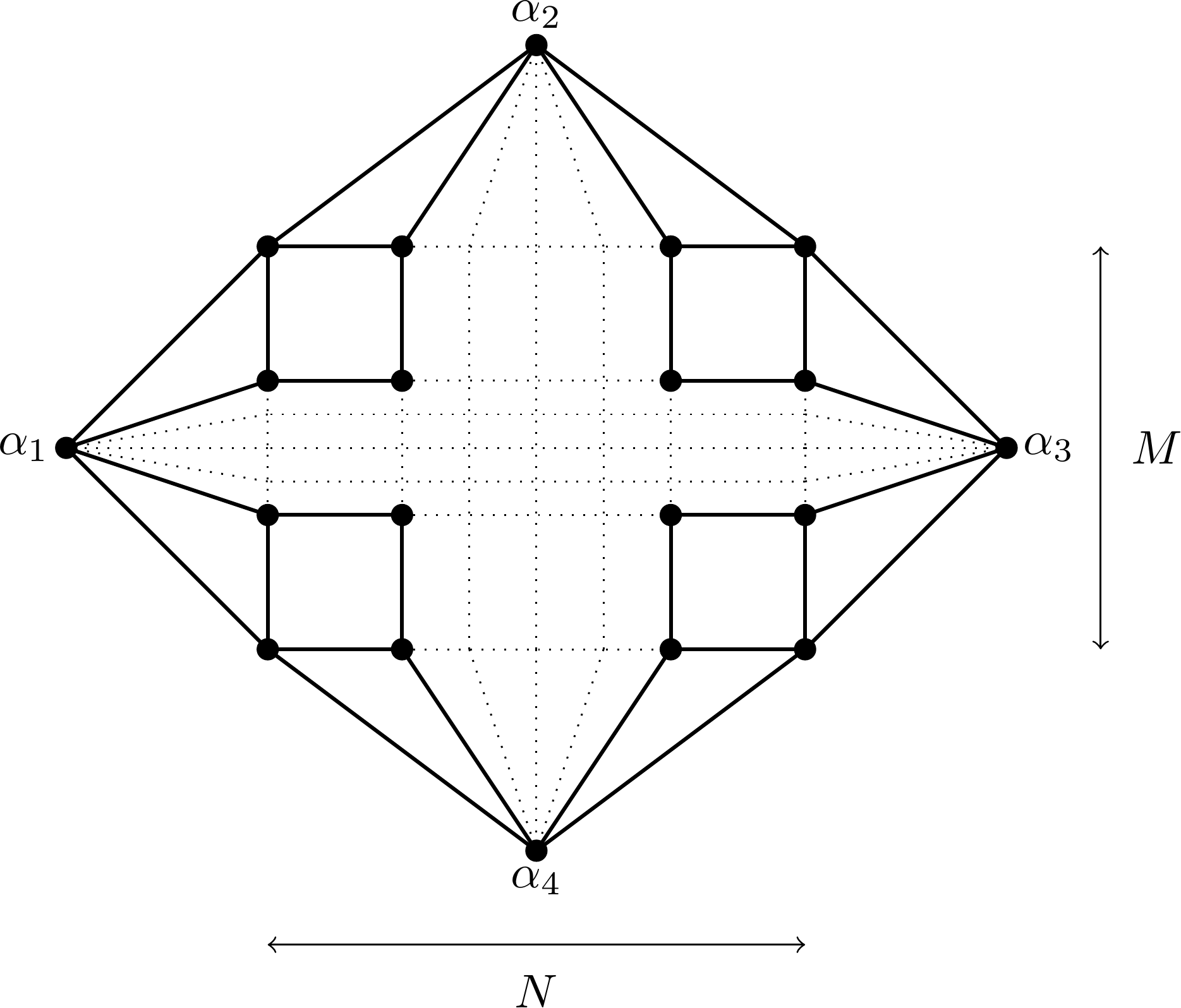}
\caption{\label{fig:4-pt-graph} The four-point graph representing the fishnet integral $I^{(D)}_{M,N}(\alpha_1,\alpha_2,\alpha_3,\alpha_4)$.
}
\end{figure}

Throughout this paper, we will consider a class of position space Feynman integrals called \emph{fishnet integrals}, cf.\ ref.~\cite{Zamolodchikov:1980mb}.
In particular, we will be interested in the four-point limits of these fishnet integrals, whose corresponding Feynman graphs are shown in figure~\ref{fig:4-pt-graph}. 
For $M=1$ (or $N=1$) we refer to them as \emph{ladder integrals}. Fishnet integrals can be defined in arbitrary space-time dimensions $D$, and the external points are labelled by points $\alpha_i\in \mathbb{R}^D$, while the internal points are labelled by $\xi_i\in \mathbb{R}^D$. An edge in the graph connecting two points labelled by $a$ and $b$ represents a propagator $[(a-b)^2]^{-D/4}$, and we integrate over all internal points. In the following, we denote the 
four-point integral obtained in this way by $I_{M,N}^{(D)}(\underline{\alpha})$, where $\underline{\alpha}=(\alpha_1,\ldots,\alpha_4)$.

Fishnet integrals are interesting for various reasons. In particular, they compute correlation functions in a
 bi-scalar fishnet theory, whose Lagrangian in $D$ dimensions reads~\cite{Gurdogan:2015csr,Kazakov:2018qbr}
 \begin{equation}
\mathcal{L}= N_c\tr \brk[s]*{ -X(-\partial_\mu \partial^{\mu})^{\frac{D}{4}} \bar{X}- Z (-\partial_\mu\partial^{\mu})^{\frac{D}{4}} \bar{Z}+\xi^2 XZ\bar{X}\bar{Z}}.
 \end{equation}
 It is easy to check that, due to the chiral nature of the interaction, the only Feynman graph contributing to the correlation function $\langle X(\alpha_1)^MZ(\alpha_2)^N\bar{X}(\alpha_3)^M\bar{Z}(\alpha_4)^N\rangle$ is the fishnet graph from figure~\ref{fig:4-pt-graph}. In a more general context it was shown that renormalisation requires additional double-trace couplings to be added to the above fishnet Lagrangian \cite{Fokken:2013aea,Sieg:2016vap}, and the beta-function has two conformal fixed points \cite{Sieg:2016vap,Grabner:2017pgm}. These additional operators are irrelevant for the correlators considered here, and so we will not discuss them any further.

Fishnet integrals enjoy an enhanced symmetry. In particular, they are invariant under conformal transformations in $D$ dimensions, where the external points carry conformal weights $D/4$, and they are annihilated by the generators of the Euclidean conformal algebra $\mathfrak{so}(1,D+1)$. This implies that we can write
\beq
I_{M,N}^{(D)}(\underline{\alpha}) = \mathcal{F}_{M,N}^{(D)}(\underline{\alpha})\,\phi_{M,N}^{(D)}(\underline{\chi})\,,
\eeq
where $ \mathcal{F}_{M,N}^{(D)}(\underline{\alpha})$ is an algebraic function carrying the conformal weight and $\phi_{M,N}^{(D)}(\underline{\chi})$ only depends on conformal cross ratios:
\beq
z_{ijkl} := \frac{\alpha_{ij}^2\alpha_{kl}^2}{\alpha_{ik}^2\alpha_{jl}^2}\,,\qquad \alpha_{ij} := \alpha_i-\alpha_j\,.
\eeq
In addition, generic fishnet integrals are invariant under the generators of the Yangian over the conformal algebra~\cite{Chicherin:2017frs,Loebbert:2020hxk} (at least in the case when the external points are not identified, i.e., no two propagators are connected to the same external point, cf. ref.~\cite{Corcoran:2021gda}). Since Yangian invariance is not directly relevant to the content of this paper, we will not dwell further on this here.
Since the four-point integrals considered in this paper depend on a single holomorphic variable, we can obtain an associated Picard-Fuchs operator by the method explicitly illustrated in section~\ref{sec:PF_fishnets}.

Finally, four-point fishnet integrals exhibit a remarkable simplicity. This was first observed in ref.~\cite{Basso:2017jwq} in $D=4$ dimensions, where it was conjectured (and recently proven in refs.~\cite{Basso:2021omx,Aprile:2023gnh}) that the integrals $I_{M,N}^{(4)}$ can be expressed as determinants of $\ell$-loop ladder integrals $I_{1,\ell}^{(4)}$. The latter are known for all values of $\ell$ in terms of classical polylogarithms~\cite{Usyukina:1992jd,Usyukina:1993ch}. A similar relation was shown to hold in $D=2$ dimensions in ref.~\cite{Derkachov:2018rot}. One of the aims of this paper is to discuss the relation between geometry and the Basso-Dixon formula in two dimensions. In the remainder of this section, we review some general results about fishnet integrals in two dimensions.


\subsection{Fishnet integrals in two dimensions}

In $D=2$ dimensions it turns out to be convenient to package the labels $\alpha_j,\xi_j\in\mathbb{R}^2$ into complex variables:
\beq
a_j := \alpha_j^1 + i\alpha_j^2 \textrm{~~~and~~~}x_j := \xi_j^1 + i\xi_j^2\,.
\eeq
It is easy to see that fishnet integrals can be cast in the form\footnote{Since from now on we only consider the case of $D=2$ dimensions, we drop the dependence of all quantities on $D$, e.g., we write $I_{M,N}$ instead of $I_{M,N}^{(2)}$, and so on.}
\beq\label{eq:I_G_complex}
I_{M,N}(\ua) = \int \left(\prod_{j=1}^\ell\frac{\rd x_j\wedge\rd \bar{x}_j}{-2i}\right)\frac{1}{\sqrt{|P_{M,N}(x,a)|^2}}\,,
\eeq
with $\ua= (a_1,\ldots,a_n)$, $\ux = (x_1,\ldots,x_\ell)$.
The integrand depends on the polynomial
\beq\label{eq:P_G_def}
P_{M,N}(\ux,\ua) = \Big[\prod_{i,j}(x_i-x_j)\Big]\,\Big[\prod_{i,j}(x_i-a_j)\Big]\,,
\eeq
where the product ranges depend on the graph topology. Up to an algebraic prefactor that carries the conformal weight, the integral then only depends on the cross ratio formed by the four complex variables:
\beq\label{eq:scaled-4-point}
I_{{M,N}}(\ua) = \big|F_{{M,N}}(\ua)\big|^2 \phi_{{M,N}}(z)\,,\qquad z := \frac{a_{23}a_{14}}{a_{21}a_{34}}\,.
\eeq

The four-point fishnet integrals have been evaluated analytically in ref.~\cite{Derkachov:2018rot} in terms of hypergeometric functions.
More precisely, we have
\beq\bsp\label{eq:fishnet_kazakov}
\big|F_{{M,N}}(\ua)\big|^2&\,=\frac{|a_{24}|^{M-N}}{|a_{14}|^{M}|a_{23}|^{M}}\frac{1}{\sqrt{|\tilde{z}|^{M}}}\,,\\
\phi_{{M,N}}(z)&\,=(2\pi)^{-M} \pi^{-M^2}\det\left[\theta_{\tilde{z}}^{j-1} \theta_{\bar{\tilde{z}}}^{k-1} \widetilde{\phi}_{{1,M+N-1}}(\tilde{z}) \right]_{1\leq j,k\leq M}\,,
\esp\eeq
with $ \theta_z := z\partial_z$ and 
\beq\label{eq:ladder_kazakov.2}
\tilde{z} =-\frac{a_{12}a_{34}}{a_{23} a_{14}}=\frac1z\, .
\eeq 
The ladder integrals $\widetilde{\phi}_{{W}}(\tilde{z}):=\widetilde{\phi}_{{1,W-1}}(\tilde{z})$ are given in terms of bilinear combinations of hypergeometric functions
\beq\label{eq:ladder_kazakov}\begin{split}
\widetilde{\phi}_{{W}}(\tilde{z}) &\,= \frac{(2\pi)^{W+1}\sqrt{|\tilde{z}|}}{(W-1)!}
 \frac{\partial}{\partial \epsilon^{W-1}}\left[\left(\frac{\Gamma\left(\frac{1}{2}-\epsilon\right)\Gamma\left(1+\epsilon\right)}{\Gamma\left(\frac{1}{2}+\epsilon\right)\Gamma\left(1-\epsilon\right)}\right)^W|\tilde{z}|^{-2\epsilon}\,\right.\\
 &\, \qquad\times\left. \bigg| {}_{W+1}F_W\left(\begin{array}{cccc}\frac{1}{2}-\epsilon &\dots &\frac{1}{2}-\epsilon &1 \\ 1-\epsilon &\dots &1-\epsilon\end{array}\Big|\tilde{z}\right)\bigg|^2\right]\Bigg|_{\epsilon= 0}
\, .
\end{split}
\eeq
The hypergeometric functions are (anti-)holomorphic at the origin, ${}_{W+1}F_W(\ldots;\tilde{z}) = 1+\mathcal{O}(\tilde{z})$, but the derivatives with respect to $\epsilon$ introduce logarithmic terms.
Equation~\eqref{eq:fishnet_kazakov} is the two-dimensional analogue of the Basso-Dixon formula of ref.~\cite{Basso:2017jwq} and was first presented in ref.~\cite{Derkachov:2018rot}.

\section{The Calabi-Yau geometries for the fishnet graphs}
\label{sec:CY_FN}
In this section, we define the Calabi-Yau (CY) geometries 
associated to integrals defined in eq.~\eqref{eq:I_G_complex}. 
We start in section \ref{subsection:geometry} by a description of the singular families of Calabi-Yau varieties defined 
by multi coverings of $\mathbb P_l=\bigtimes_{i=1}^l \mathbb{P}_i^1$ branched at eq.~\eqref{eq:P_G_def}, so that the integral becomes a real 
bilinear of the periods of the Calabi-Yau variety. The periods of the Calabi-Yau varieties associated to the ladder integrals  play a special 
role: we will argue that the periods of more general fishnet geometries can be obtained 
by anti-symmetric powers of periods of ladder geometries. 
Moreover, to the latter we can associate generically smooth families of Calabi-Yau manifolds $\mathcal{M}_{1,N}$, as well as smooth mirror families $\mathcal{W}_{1,N}$. Using a deep result of Beukers and Heckman  on the geometric monodromy group action ${\cal G}_N$ on periods of $\mathcal{M}_{1,N}$ we can 
understand the periods of general fishnet geometries and their degenerations by geometric representation theory of $\cG_N$. We provide 
the necessary information about these building blocks in section \ref{sec:laddergeometries}.   Finally, in section~\ref{sec:defvolumes}, we discuss two different interpretations of fishnet integrals as volumes associated to CY manifolds.

\subsection{Calabi-Yau varieties associated to ladder integrals in two dimensions} 
\label{subsection:geometry}
In ref.~\cite{Duhr:2022pch} we argued that fishnet integrals in $D=2$ dimensions are closely related to Calabi-Yau (CY) periods on Calabi-Yau varieties. Roughly speaking, a Calabi-Yau manifold $\mathcal{M}$ is a K\"ahler manifold with trivial canonical bundle, $K_\mathcal{M}=0$ (see ref.~\cite{MR1963559} for the precise definition we use in this paper). CY varieties $\mathcal{M}$ include more generally the singular CY geometries in complex deformation families of CY manifolds. As in ref.~\cite{Bonisch:2021yfw}, the restricted physical parameters of Feynman integrals identify them also in the application here with Calabi-Yau periods on 
these singular loci. Similar to ref.~\cite{Bonisch:2021yfw}, our strategy is again to find the smooth family and to argue that the 
restriction to the singular loci defines a suitable CY motive. For the ladder integrals considered here, the smooth motives all
turn out to be hypergeometric (Gel\cprime fand, Kapranov and Zelevinsk\u{\i}) motives associated to complete intersections, cf.\ refs.~\cite{Klemm:2019dbm,Bonisch:2020qmm}. 
 
The relevant geometries $\mathcal{M}$ are $d$-fold coverings over the base  $\mathbb P_l$ defined by an equation of the form
\beq\label{eq:fishnet_model}
y^d=P(\ux,\ua)\,.
\eeq
The coverings are finitely branched at $B=\{P(\ux,\ua)=0\}$. For now, we keep the discussion general, and we consider fishnet graphs with an arbitrary  number of external points $\underline{a}$, and we specialise below to the four-point graphs in figure~\ref{fig:4-pt-graph}. The coverings defined by eq.~\eqref{eq:fishnet_model} are projective and therefore K\"ahler. By the adjunction formula, the canonical class of $K_{\mathcal M}$ is trivial if 
\beq 
\frac{d}{d-1}K_{\mathbb P_l}= \frac{d}{d-1} \bigoplus_{i=1}^\ell 2 H_i =[B]=\nu \bigoplus_{i=1}^\ell H_i \ .
\label{eq:CYcond}
\eeq
Here $H_i=\{x_i=0\}$ is the hyperplane class of the $i^{\textrm{th}}$ $\mathbb{P}^1$, and $\nu\in \mathbb{N}$ 
is the common degree of $P(\ux,\ua)$ in each $x_i$. The triviality of $K_{\mathcal M}$ implies the existence of 
a unique, nowhere vanishing $(\ell, 0)$-form. As a section of $K_\mathcal{M}$, it is given in the homogeneous coordinates $[x_i:w_i]$ of each $\mathbb{P}^1$ by
\beq\label{eq:Omega}
\Omega(\uz)= \frac{\bigwedge_{i=1}^{\ell} (w_i\,  \rd x_i-x_i\, \rd w_i)}{P(\ux:\uw,\uz)^\frac{d-1}{d}}=
\frac{1}{F(\ua)} \frac{\bigwedge_{i=1}^{\ell} (w_i\, \rd x_i-x_i\, \rd w_i)}{P(\ux:\uw,\ua)^\frac{d-1}{d}}
\ .
\eeq
Here we have to make a comment. The polynomial $P(\underline{x},\underline{a})$ depends on the external points $\underline{a}$. Conformal invariance, however, implies that the only non-trivial functional dependence can be in cross ratios $\underline{z}$, up to an algebraic function $F(\underline{a})$ that carries the conformal weight (cf.~eq.~\eqref{eq:scaled-4-point}). Correspondingly, we can define the $(\ell, 0)$-form $\Omega(\uz)$ such that it only depends on cross ratios. The function $F(\ua)\in {\overline {\mathbb{Q}(\ua)}}$ lives in a simple algebraic extension of $\mathbb{Q}(\ua)$ completely fixed by our normalisation $\Omega(\uz)$, chosen here as in Definition 4.3.\ of ref.~\cite{MR1963559}:
\beq
\label{eq:Volume}
\phi(\uz)=e^{-K(\uz)}=\int_\mathcal{M} (-1)^{\ell(\ell-1)/2} \left(\frac{i}{2}\right)^\ell \Omega(\uz) \wedge \overline{\Omega}({\uz}) \, ,
\eeq
where $\mathcal M_{\uz}$ denotes the fiber over the point $\uz$, $\omega$ denotes the K\"ahler form on $\mathcal M_{\uz}$, and $K(\uz)$ is the K\"ahler potential on the complex moduli space $\mathcal{M}_{\textrm{cs}}$ parametrised by $\uz$.
Note that $\phi(\uz)$ has two interpretations as a volume, as we will
further explain in section \ref{sec:defvolumes}. 
On a family of CY $\ell$-folds parametrised by $\uz$,
we can calculate the monodromy-invariant real quantity $\phi(\uz)$ in terms of the periods (see also ref.~\cite{Duhr:2023bku})
\beq
\label{eq:periodpairing} 
\phi(\uz)=(-i)^{\ell^2} {\underline \Pi}(\uz)^\dagger \Sigma {\underline \Pi}(\uz) \ .
\eeq
Here $\underline{\Pi}(\uz)$ is a vector of period integrals of $\Omega(\uz)$, 
\beq\label{eq:period_vector}
\underline\Pi(\uz) = 2^{-\frac{\ell}{2}}\Big(\int_{\Gamma_0}\!\!\!\Omega(\uz),\hdots,\int_{\Gamma_{b_\ell-1}}\!\!\!\!\!\!\Omega(\uz)\Big)\,,\,\,\,\,\, b_\ell = \dim H_\ell(\mathcal{M},\mathbb{Z})\,,
\eeq
 where the $\Gamma_j$ label an integral basis of the middle homology $H_\ell(X, \mathbb{Z})$, 
and $\Sigma$ is the corresponding integral intersection form, which is even for $\ell$ even 
and odd for $\ell$ odd. In particular, for $\ell$ odd, it can always be chosen to be symplectic. A period 
of the family $\mathcal M_{\uz}$ is a solution of the \emph{Calabi-Yau Picard-Fuchs ideal}, denoted in the following by $\CPFI(\mathcal M_{\uz})$. The existence of the $\CPFI(\mathcal M_{\uz})$ is equivalent to the flatness of the Gauss-Manin connection on the Hodge 
bundle over the complex moduli space $\mathcal{M}_{\textrm{cs}}$.
Every solution vector $\underline{\Pi}(\uz)$ fulfils a set of quadratic relations, which follow from Griffiths transversality of the Gauss-Manin connection (see ref.~\cite{Bonisch:2021yfw} for a review from the
Feynman graph perspective): 
\begin{equation} {\underline \Pi}(\uz)^T \Sigma  \partial_{z_{I_1}}\ldots\partial_{z_{I_k}}  {\underline \Pi} =\left\{
\begin{array}{ll} 0\ , & \ \ {\rm if} \ k < \ell\,,\\[2 mm] 
C_{I}(z) \in \mathbb{Q}(\uz)\ , & \ \ {\rm if} \  k=\ell \ . \end{array}\right.  
\label{Griffiths}
\end{equation}

So far the discussion applies to very general families defined by $d$-fold branched coverings as in eq.~\eqref{eq:fishnet_model}. We now focus on the special case we are interested in, namely the fishnet graphs from figure~\ref{fig:4-pt-graph}. In that case, there is a single cross ratio, and so we write simply $z$ instead of $\uz$. In particular, the choice $(d,\nu)=(2,4)$ satisfies the condition $\frac{2 d}{d-1}=\nu$ implied by eq.~\eqref{eq:CYcond}.\footnote{The multi-parameter cases and the choice $(d,\nu)=(3,3)$ will
be discussed in an accompanying paper~\cite{CY_paper_2}.}  
Writing $P=P_{M,N}$, as defined in eq.~\eqref{eq:P_G_def}, we arrive at the singular fishnet geometry $\mathcal{M}_{M,N}$:
\beq
y^2=P_{M,N}(\ux,z) \,.
\label{eq:Fishnetgeometry} 
\eeq
From now on, we will indicate quantities related to $\mathcal{M}_{M,N}$ by an index $(M,N)$, e.g., we will write $F_{M,N}(\ua)$, $\Omega_{M,N}(z)$, $\uPi_{M,N}(z)$ and $\phi_{M,N}(z)$ instead of $F(\ua)$, $\Omega(z)$, $\uPi(z)$ and $\phi(z)$. 
Moreover, $z$ will denote the cross ratio defined as in eq.~\eqref{eq:scaled-4-point}. 
From eq.~\eqref{eq:periodpairing} we then see that the value of the fishnet integral $\phi_{M,N}(z)$ can be computed from the periods ${\underline \Pi}_{M,N}(z)$ and the intersection form $\Sigma_{M,N}$ of $\mathcal{M}_{M,N}$:
\beq
\label{eq:PiSPi} 
\phi_{M,N}(z)=(-i)^{\ell^2} {\underline \Pi}_{M,N}(z)^\dagger \Sigma_{M,N} {\underline \Pi}_{M,N}(z) \,.
\eeq
To each fishnet integral we can associate the Calabi-Yau 
variety $\mathcal{M}$ defined by eq.~\eqref{eq:Fishnetgeometry} or in case of the ladder integrals alternatively 
a Calabi-Yau manifold  as in eq.~\eqref{eq:geomrealisation}. 
The geometrisation implies that after absorbing the factor  $|F(\ua)|^2$ 
we can geometrise the integral in terms of a real period bilinear in eq.~\eqref{eq:periodpairing} 
that has geometric interpretations in terms of two different volumes of a
Calabi-Yau manifold, as we explain in detail in section \ref{sec:defvolumes}.

\subsection{The Calabi-Yau manifolds associated to the ladder integrals}
 \label{sec:laddergeometries}
 Let us start  from the  Picard-Fuchs equation for the $N$-loop ladder integral. The latter can be derived by the method outlined in section \ref{sec:PF_fishnets} from the general form of the period integral in eq.~\eqref{eq:period_vector} with $\Omega(\uz)$ given in eq.~\eqref{eq:Omega}.
 We note that for the ladder integrals the loop order $\ell$, 
 the complex dimension ${\rm dim}_\mathbb{C}({\cal M}_{1,N})$ and $N$ are equal
 \beq
 \ell= N={\rm dim}_\mathbb{C}({\cal M}_{1,N})\,, 
 \eeq
so that below we can use $\ell$, $d$ or $N$ as appropriate for the context.  
 We find that  the associated  motive is of Gel\cprime fand, Kapranov and Zelevinsk\u{\i} type before the one-parameter specialisation, and in the one-parameter limit (where the integral only depends on the cross ratio $z$), it corresponds to the hypergeometric system (see ref.~\cite{MR0974906} for the conventions):
\beq 
\label{eq:hypergeometricmotive}
L_N \coloneqq L_{1,N}=\theta_z^{N+1} - z \left(\theta_z+\frac{1}{2}\right)^{N+1}\,,
\eeq
with the Riemann $\mathcal P$-Symbol 
\begin{equation}
\begin{aligned}
	\mathcal P	&	\begin{Bmatrix*}[l]	%
							z 				& 1 & \infty			\\ \hline
							0 &0 &\frac{1}{2}	\\
                            0 &1 &\frac{1}{2}	\\
                            \vdots  &\vdots &\vdots\\
                             0 &N-1 &\frac{1}{2}	\\
                             0 &\frac{N+1}{2}-1 &\frac{1}{2}	\\ 
        \end{Bmatrix*}~.
\label{Riemannepseven}
\end{aligned}
\end{equation}
This one parameter hypergeometric motive has been geometrically realised in ref.~\cite{MR1201748} as the $\CPFI$ for the mirror $\mathcal{M}_{1,N}$ of the complete intersection of $N+1$ quadrics in $\mathbb{P}^{2N+1}$, abbreviated as 
\beq
\label{eq:geomrealisation}
\mathcal{W}_{1,N}=\left( \mathbb{P}^{2N+1} \| 2,\ldots, 2\right) \,. 
\eeq
Mirror symmetry is a well studied symmetry relating the $h^{1,1}({\cal W})$ complexified K\"ahler structure deformations of a CY $N$-fold ${\cal W}$ to the $h^{1,N-1}({\cal M})$ complex structure deformations of the mirror CY 
$N$-fold  ${\cal M}$ and vice versa, see ref.~\cite{Bonisch:2021yfw} for a 
review from the Feynman intregral perspective and more references. 
Therefore the simple one complex structure parameter families of CY $N$-folds have 
been first studied as mirrors of simple complete intersections having only one K\"ahler structure deformation as in \eqref{eq:geomrealisation}. In this case the mirror can be constructed as a resolution of an orbifold with respect to the action of a discrete group $\Gamma\subseteq {\rm SU}(N)$ on $\mathcal{W}_{1,N}$, i.e., it is defined 
as $\mathcal{M}_{1,N}=\widehat{\mathcal{W}_{1,N}/\Gamma}$. The resolution of $\mathcal{M}_{1,3}=\widehat{\mathcal{W}_{1,3}/\mathbb{Z}_8^3}$ 
has been constructed in ref.~\cite{Bonisch:2022mgw} with methods that
apply in general. This realises the Calabi-Yau motive as the motive of a smooth Calabi-Yau $N$-fold $\mathcal{M}_{1,N}$.

The differential operator $L_N$ has a point of maximal unipotent monodromy (MUM) at $z=0$. At such a MUM point the local index, here $0$, is $(N+1)$-times degenerate. That implies that one of the solutions has a $\log^Nz$ behaviour and the monodromy $M$ around $z=0$ is maximal unipotent, i.e. $(M-{\bf 1})^k$ vanishes only for $k\ge N+1$. In fact, the operator $L_{N}$ is a CY operator of degree $N+1$ (cf.\ section~\ref{sec:CY_ops}). The fundamental solution that is holomorphic at the MUM-point $z=0$ is:
\beq
\label{eq:ladder_holomorphic} 
\Pi_{N,0}(z)  \coloneqq \Pi_{1,N,0}(z)  = \sum_{i=0}^\infty \binom{-\frac{1}{2}}{\ i}^{N+1}\,z^i = {}_{N+1}F_N\left({ \frac{1}{2},\ldots,\frac{1}{2}};1,\ldots,1;{z}\right)\ .
\eeq
Note that $b_N(\mathcal{M}_{1,N})=N+1$  and the Hodge numbers are:
\beq 
\label{eq:1111}
h^{N-k,k}(\mathcal{M}_{1,N})=1, \quad  k=0,\ldots, N\ .
\eeq
We call this a \emph{Hodge structure of type $(1,1,\ldots,1,1)$}. 
Let $J_n$ denote the rank $n$ exchange matrix $(J_n)_{ij}=\delta_{i,n-j}$. We choose  
the rank $N+1$ intersection matrix of $\mathcal{M}_{1,N}$ as 
\beq
\label{eq:Sigmal}
\Sigma_N=\left(\begin{array}{lr} {\bf 0}&J_{\frac{N+1}{2}}\\ -J_{\frac{N+1}{2}}&{\bf 0}\end{array}\right), \quad
N \ {\rm odd}, \qquad\quad \Sigma_N=\left(\begin{array}{lcr} {\bf 0}&0&J_{\frac{N}{2}}\\
0 & m & 0\\ J_{\frac{N}{2}}&0&{\bf 0}\end{array}\right), \quad N \ {\rm even} \ .
\eeq
This choice is made such that at a given point of \emph{maximal unipotent monodromy} or \emph{MUM-point} (see section~\ref{sec:diffopsinone} for details) it induces a pairing 
between those cycles whose periods degenerate like $z^\lambda\log^{N-k}z$ and  $z^\lambda\log^{k}z$ for  $k=0,\ldots, d$, respectively. 

The Zariski closure ${\overline{\cal G}_N}$ of the geometric monodromy group\footnote{This closure is equivalent to the Hodge group of the motive.} ${\cal G}_N$ is irreducible over $\mathbb{C}$ and given by 
\beq \label{eq:Hodgegroups}
{\overline { \cal G}_N} =\left\{\begin{array}{ll}  {\rm Sp}(N+1) & \quad N \  {\rm odd},\\
  {\rm SO}(N/2,N/2 +1) & \quad N \ {\rm even} \end{array}\right. . 
\eeq
This statement was proven for the hypergeometric motive in
eq.~\eqref{eq:hypergeometricmotive} in  Theorem 6.5 of ref.~\cite{MR0974906} for many more hypergeometric systems, together with the field of definition $F$ that occurs in the monodromy actions for $m=1$. Let us note that, in order to evaluate the bilinear in eq.~\eqref{eq:periodpairing}, it is not mandatory that we have fixed a basis of periods over $\mathbb{Z}$. 
A $\cG_N(\mathbb{R})$  basis change with $\overline{\cG}_N$ as in eq.~\eqref{eq:Hodgegroups} is obviously 
admissible. The geometric monodromies act on the space of solutions, and in 
particular the vector spaces of the solution to eq.~\eqref{eq:hypergeometricmotive} form by eq.~\eqref{eq:Hodgegroups} irreducible fundamental representations of the corresponding group ${\overline{\cal G}_N}$. We call these irreducible representations $\mV$. Representation theory therefore governs the monodromies and the Hodge structures associated to  the symmetric and antisymmetric powers discussed in section \ref{eq:opsonops}.       

For $N$ odd, we can make the statement to hold over $\mathbb{Z}$, 
and for $N$ even with a  choice of $m$. The $\widehat \Gamma$-class formalism described  in ref.~\cite{Bonisch:2020qmm} promotes
the Frobenius basis in eq.~\eqref{eq:Frobenius} at the MUM-point $z=0$ to an integral basis in eq.~\eqref{eq:period_vector}, provided that the intersection numbers between the restriction $H$ 
of the hyperplane class from $\mathbb{P}^{2N+1}$ to $\mathcal{W}_{1,N}$ and the Chern 
classes of the tangent bundle are known.  This latter data is defined by
\beq
\label{eq:intersections}
c_m(\mathcal{W}_{1,N})=\left(\left.\frac{(1+x)^{2N+1}}{(1+2 x)^{N+1}} \right|_m \right) H^m, \qquad 
H^N=2^{N+1}\ , 
\eeq
where $|_m$ means taking the $m^{\textrm{th}}$ coefficient. 
For the geometrical realisation $\widehat{\mathcal{M}_{1,N}}$, we checked that eq.~\eqref{eq:periodpairing} equals eq.~\eqref{eq:ladder_kazakov}. This gives the coefficients of the $\epsilon$ expansion of eq.~\eqref{eq:ladder_kazakov} an interpretation in terms of intersection numbers of $\mathcal{W}_{1,N}$. For example, the Euler numbers of $\chi(\mathcal{W}_{1,N})=0,24,-128,960,-6912,51051,\ldots $, for $N=1,2,3,4,5,6,\ldots$ that follow from eq.~\eqref{eq:intersections} describe an admixture of the fundamental solution to 
the highest $\log^Nz$ logarithmic solution of the form $\frac{\zeta_N}{(2 \pi i)^N} \chi(\mathcal{W}_{1,N})$ etc.

 Since the periods are solutions of the $\CPFI$, we review in section~\ref{sec:DEQ_ops} general results about $\CPFI$s for one-parameter families of CY varieties.
 
 \subsection{Fishnet integrals as Calabi-Yau volumes}
\label{sec:defvolumes}
There are two relations of the fishnet integrals to \emph{volumes} of Calabi-Yau  varieties, which are both conceptually very interesting and related by mirror symmetry. 
Let $\mathcal{M}_s$ be the \emph{singular} Calabi-Yau variety, 
which we associated to the fishnet integrals in section~\ref{subsection:geometry}. The definition of the volumes  
is most easily stated if we have a smooth Calabi-Yau realisation $\mathcal{M}$ and a smooth mirror $\mathcal{W}$ of it, even though there will be generalisations for generically singular families.  For example, for the ladder diagrams the smooth $\mathcal{W}$ is given in \ref{eq:geomrealisation}, while a
\emph{smooth} Calabi-Yau manifold $\mathcal{M}$ was defined as the resolved orbifold $\mathcal{M}_{1,N}$. Let us assume we have such 
smooth models. As emphasised in ref.~\cite{Duhr:2022pch} the fishnet integrals defined from periods on $\mathcal{M}$ compute the \emph{quantum volume} of the mirror Calabi-Yau variety $\mathcal{W}$. A second interpretation is as the calibrated volume of $\mathcal{M}$ itself.
We will  give  short accounts of booth concepts. 

\paragraph{The fishnet integral  as the calibrated volume of $\mathcal{M}$}
Let us go back to the definition of $\phi(\uz)$ in eq.~\eqref{eq:periodpairing}, which up to the
absolute value of the factor $F(\ua)\in \overline{\mathbb{Q}(\ua)}$
is the fishnet integral and encodes its non-trivial dependence 
on the transcendental periods, which are generalisations of rational or elliptic functions. As explained in~\cite{MR1963559}, see again Definition 4.3.,
there is a \emph{canonical calibration} given by
\beq
\label{eq:calibratedVolume}
\phi(\uz)=\int_{\mathcal{M}} (-1)^{\ell(\ell-1)/2} \left(\frac{i}{2}\right)^\ell \Omega(\uz) \wedge \overline{\Omega}({\uz})
=\int_{\mathcal{M}} \frac{\omega^\ell}{\ell !} ={\rm Vol}(\mathcal{M}) \, ,
\eeq
which relates $\Omega(\uz)$ to the K\"ahler form $\omega$. 
The volume form of $\mathcal{M}$ is defined by
\beq
\label{eq:vol_cl_def} 
\textrm{vol} \coloneqq \frac{\omega^{\ell}}{\ell!}\, .
\eeq
Then, by integrating the latter over $\mathcal{M}$, we obtain  the 
\emph{calibrated volume} of $\mathcal{M}$.
The calibrated volume is real, positive and monodromy invariant. In contrast to the quantum volume in eq.~\eqref{eq:Vol_q_def} it can be defined classically, given the K\"ahler form $\omega$ defined by the calibration.
We related $\phi(\uz)$ to the  \emph{calibrated volume} of $\mathcal{M}$. However, its \emph{definition}  depends by eq.~\eqref{eq:calibratedVolume} on the physical parameters encoded in $\uz$.

\paragraph{The fishnet integral as the quantum volume of $\mathcal{W}$}
In ref.~\cite{Duhr:2022pch} we used mirror symmetry and proposed another 
relation to a volume whose \emph{value} depends on the physical 
parameters encoded in $\uz$, while its \emph{definition} stays fixed.
This is the \emph{quantum volume} of the mirror $\mathcal{W}$. To define it, 
we define a K\"ahler class on $\mathcal{W}$  
\beq
\omega = \sum_{i=1}^{d_{\mathcal{M}}} t_{i}^{\mathbb{R}}(\uz)\,\omega^{(i)}
\eeq
using a fixed basis $\big\{\omega^{(i)}\big\}_{i=1,\ldots,d_{\mathcal M}}$ of $H^{1,1}(\mathcal{W})$. 
At a MUM-point at $\uz=0$, which, e.g., the ladder 
Calabi-Yau manifolds have, we have a relation  $t_{i}^{\mathbb{R}}(\uz) = \textrm{Im}\,t_i(\uz)$ given  by the \emph{mirror map}, i.e., in terms of the periods (cf.~eq.~\eqref{eq:mirror_map_CY_op})
\beq\label{eq:mirror_map}
t_{i}(\uz) = \frac{1}{2\pi i}\,\frac{{\Pi}_{i}(\uz)}{{\Pi}_{0}(\uz)}\,,\quad i=1,\ldots,d_{\mathcal M}\,.
\eeq

We can then use eq.~\eqref{eq:vol_cl_def} to obtain the classical volume of the mirror $\mathcal{W}$:
\beq\bsp\label{volcl}
\textrm{Vol}_{\textrm{cl}}(\mathcal{W}) &\,= \int_{\mathcal{W}}\frac{\omega^\ell}{\ell!} = \frac{1}{\ell!}\sum_{i_1,\cdots,i_{\ell}}C^{\textrm{cl}}_{i_1,\cdots,i_{\ell}}\,t_{i_1}^{\mathbb{R}}(\uz)\cdots t_{i_\ell}^{\mathbb{R}}(\uz)\,,
\esp\eeq
where the $C^{\textrm{cl}}_{i_1,\cdots,i_{\ell}}$ are explicitly-computable integers, namely the (classical) intersection numbers of $\mathcal{M}$, which for the ladder geometries are defined by 
eq.~\eqref{eq:intersections}, i.e. here we have only the intersection number of $\ell$ divisors given by $C^{\textrm{cl}}_{H\cdots H}=2^{\ell +1}$. 
While its definition is fixed, it depends on the values of $\uz$ via 
\eqref{eq:mirror_map}. Moreover, it gets quantum corrected by world-sheet instanton effects to the quantum volume ($q_i=\exp(2 \pi i t_i)$):
\beq
\textrm{Vol}_{\textrm{q}}(\mathcal{W}) = \textrm{Vol}_{\textrm{cl}}(\mathcal{W}) + \mathcal{O}\big(e^{-t_i^{\mathbb{R}}(\uz)}\big)\,,\qquad \textrm{as }\uz\to0\,.
\label{eq:Vol_q_expanded}
\eeq
It is the key observation of mirror symmetry that the latter 
can be encoded in
\beq\label{eq:Vol_q_def}
\textrm{Vol}_{\textrm{q}}(\mathcal{W}) = (-i)^{{\rm dim}^2(\mathcal{W})}\frac{\uPi(\uz(\uq))^{\dagger}\Sigma\uPi(\uz(\uq))}{\big|\Pi_{0}(\uz(\uq))\big|^2}\, ,
\eeq  
where $\underline{\Pi}$ are the periods on $\mathcal{M}$ and the 
relation $\uz(\uq)$ is given by inverting eq.~\eqref{eq:mirror_map}.
This quantity  is also real and positive, but not quite monodromy invariant, because of the normalisation by $|\Pi_0(\uz(\uq))|^2$  and the fact that $\Pi_0(\uz)$ undergoes monodromy changes. This normalisation and the absence of instanton corrections in low dimensions ${\rm dim}^2(\mathcal{W})=1,2$ is discussed in ref.~\cite{Duhr:2022pch}.
Restoring the monodromy invariance, we  write:
\beq
\label{eq:phivolume}
\phi(z(q)) = \big|\Pi_0(\uz(q))\big|^2\,\textrm{Vol}_{\textrm{q}}(\mathcal{W})\,.
\eeq
That is,  the fishnet integrals are proportional to the quantum 
volume of the associated mirror manifolds $\mathcal{W}$. 

In this section we have given two interpretations of eq.~\eqref{eq:periodpairing} in terms of Calabi-Yau volumes that  apply directly to the ladder graphs where the geometry is smooth. For the general fishnet graphs, we can take eqs.~\eqref{eq:phivolume} and \eqref{eq:PiSPi} as the definition of the volume of the corresponding singular varieties. The calculation of these volumes by iterated integrals in the general case will be discussed in section \ref{sec:volumes}.

%% file: 4-pt.tex

\section{Picard-Fuchs operators for one-parameter families of Calabi-Yau varieties}
\label{sec:DEQ_ops}


\subsection{Differential operators in one variable}
\label{sec:diffopsinone}
In this section, we briefly review some mathematical concepts related to differential operators in one variable. We follow the review~\cite{Dmodulebook} and 
consider a differential operator of degree $n$ over $\mathbb{P}^1$:
\beq\label{eq:L_genericpartial]}
L = \partial_z^n + a_{n-1}\partial_z^{n-1}  +\ldots +  a_1\partial_z + a_0, \  {\rm with} \ a_i\in \mathbb{C}(z)\ .   
\eeq
Its $n$-dimensional $\mathbb{C}$-vector space of solutions is
\beq
\Sol(L) = \{f:\mathbb{C}\to \mathbb{C}:Lf=0\}\,.
\eeq
If $\Delta_L(z)\in \mathbb{C}[z]$ is the least common multiple of the denominators  
of the $a_i(z)$, then the solutions  can have singularities only at the  components of the \emph{discriminant divisor} $z=z_i$ with $z_i$ given by the roots of $\Delta_L(z)=0$, possibly supplemented by $z=\infty$ in $\mathbb{P}^1$. A one-parameter $\CPFI(\mathcal{M}_{z})$ is
generated by a single differential operator $L$ of the form \eqref{eq:L_genericpartial]}. 

If the $\CPFI(\mathcal{M}_{z})$ has the properties in eqs.~\eqref{eq:1111} and~\eqref{eq:Sigmal}, then $n= {\rm dim}_{\mathbb{C}}(\mathcal{M})+1$, and ${\underline \Pi}(z)$ spans  
$\Sol(L)$. A short calculation reveals that eq.~\eqref{Griffiths} implies (with $a_{n}=1$):
\beq
\label{eq:selfadjoint}
L^* \alpha(z)=(-1)^{n} \alpha(z) L, \qquad  {\rm with }\ L^*=\sum_{i=0}^n (-\partial_z)^i a_i(z)\ ,
\eeq
where $\alpha(z)= c \, C_I(z) = \kappa \exp\left(-\int_0^z \frac{2}{n} a_{n-1}(z')d z'\right)$.\footnote{$C_I$ is canonically normalized by $\kappa$, which is the intersection
 number on the mirror CY.} 
Operators that satisfy the property in eq.~\eqref{eq:selfadjoint} are called 
\emph{essentially self-adjoint}. 

A local basis for $\Sol(L)$ can be constructed using the Frobenius method.
We use the coordinates $\delta_i=z-z_i$ and $\delta_\infty=1/z$, and we 
transform the operator to logarithmic derivatives $\theta_*=\delta_* \frac{d}{d \delta_*}$:
\beq
\label{eq:L_standard}
L(\theta_*,\delta_*) =\sum_{k=0}^n p_k(\delta_*) \theta^k_*\,,   {\rm ~~~with~~~} \ p_k(\delta_*)\in \mathbb{C}[\delta_*]\ , \eeq
where the $p_k(\delta_*)$ have no common factor. If we view the $L(\theta_*,\delta_*)$ as polynomials in the formal variables $\theta_*$ and $\delta_*$, then the \emph{local exponents}\footnote{In some older standard references on linear differential equations like ref.~\cite{MR0010757} they are called \emph{indicials}.} at $z_*$ 
are given by the roots of the polynomial $L(\theta_*,\delta_*)|_{\delta_*=0}$, which is of degree $n$
in $\theta_*$. If these  roots $\{\lambda\}$ are $n$-fold, i.e.\ maximally degenerate at $z_*$, the latter point is
called a \emph{point of maximal unipotent monodromy} or MUM-point.
Let us assume that at a singularity,
which we choose to be at $z=0$ to simplify notation, we have $k$ 
distinct roots $\lambda_i$ with multiplicity $m_i$ (with $\sum_{i=1}^k m_i=n$). 
For simplicity, we first assume that $\lambda_i-\lambda_j\notin \mathbb{Z}$.
Then the  Frobenius method guarantees that, for each $\lambda_i$, one can construct 
$m_i$ independent solutions of the form 
\beq
\label{eq:Frobenius}
y^{(i,r)}(z) = z^{\lambda_i}\sum_{k=0}^{r} \varpi^{(i)}_{r-k}(z) \frac{1}{k!}\log^k z , \qquad r=0,\ldots,m_i-1\, .
\eeq
these soltuions form the \emph{Frobenius basis} of $\Sol(L)$ in which the local monodromy is lower triangular. 
We can fix the  $m_i$ holomorphic power series in $z$ uniquely by demanding $\varpi^{(i)}_0(0)=1$ and $\varpi^{(i)}_k(0)=0$, $k=1,\ldots,m_i-1$. In particular, at a MUM-point the highest logarithmic degeneration is $z^{\lambda_i}\log^{n-1} z$. 
If the differences of $k$ indices $\lambda_{j_s}$, $s=1,\ldots,k$ are in $\mathbb{Z}$, the highest logarithmic degeneracy is ${\rm max}\{m_{j_s}-1|s=1,\ldots,k\}$, but more logarithmic 
solutions can appear, as the logarithmic degeneracies of the larger indices can be shifted. One still gets $\sum_s m_{j_s}$ independent solutions involving $\sum_s m_{j_s}$ power series, but the latter 
need to be indexed differently.

The solutions of the differential operator $L$ generating a $\CPFI(\mathcal{M}_{z})$ are the periods of~$\mathcal{M}_z$. Besides eq.~\eqref{Griffiths}, they 
obey further restrictions. Part of them are valid for periods on any algebraic variety 
of dimension $\ell$. For example, the non-logarithmic monodromies are finite, and so all local exponents must be rational, $\lambda_k\in\mathbb{Q}$. Moreover, Landmann's theorem~\cite{MR344248} implies that the highest logarithmic degeneration at singular locii is $\sim \delta_*^{\lambda_i} \log^\ell\delta_*$. The latter implies that a differential operator $L$ describing periods can only have a MUM-point if its degree is $n=\ell+1$.
Others are specific for period degenerations of a $\CPFI(\mathcal{M}_{z})$, see ref.~\cite{Bonisch:2020qmm}
for examples and references. 

\subsection{Calabi-Yau operators}
\label{sec:CY_ops}
In this section, we review a class of differential operators in one variable that play a prominent role in the theory of one-parameter families of CY $\ell$-folds. Before we state the definition, we need to introduce some concepts.

Throughout this section we assume that $L$ is a differential operator of degree $\ell+1$ with only regular-singular points. We assume that $L$ has a MUM-point at $z=0$, and admits a basis of solutions of the form
\beq\label{eq:CY_op_basis}
y_k(z) = y_0(z)\frac{1}{k!}\log^kz + \mathcal{O}(z)\,,\qquad 0\le k\le \ell\,.
\eeq
The \emph{mirror map} is defined by
\beq\label{eq:mirror_map_CY_op}
t(z) = \frac 1{2\pi i}\frac{y_1(z)}{y_0(z)} = \frac{\log z}{2\pi i} + \mathcal{O}(z)\,.
\eeq
Its exponential is a holomorphic function of $z$:
\beq
q(z) = e^{2\pi i\,  t(z)} = z+\mathcal{O}(z^2)\,.
\eeq

The periods can be used to define another set of functions that are holomorphic in a neighbourhood of $q=0$. They are defined as follows: Let $\alpha_m(z) = u_{m,m}(z)^{-1}$, where the $u_{m,k}(z)$ are determined recursively:
\beq\label{eq:u_rec}
u_{m,k}(z) = \theta_z\left[\frac{u_{m-1,k}(z)}{u_{m-1,m-1}(z)}\right] \textrm{~~~and~~~} u_{0,k}(z) = y_{k}(z)\,.
\eeq
The functions $\alpha_m(z)$ are holomorphic at $z=0$, and called the \emph{structure series of $L$}. In appendix~\ref{app:proof_IIQ} we present a formula that allows one to compute the structure series as a ratio of determinants of periods.

We define an \emph{almost Calabi-Yau operator} $L$ of degree $\ell+1$ as a differential operator with regular-singular singularities   such
that:
\begin{itemize}
\item $(i)$ $L$ is essentially self-adjoint,
\item $(ii)$ $L$ has a MUM point at $z=0$, and there is a local basis of solutions as in eq.~\eqref{eq:CY_op_basis}.
\end{itemize}
In refs. \cite{Almkvist1,Almkvist2,BognerCY,BognerThesis,MR3822913} 
a \emph{Calabi-Yau operator} is required to have the additional property that
\begin{itemize}
\item $(iii)$ the holomorphic functions $y_0(z)$, $q(z)$ and $\alpha_m(z)$ are $N$-integral, by which we mean that there is an integer $N$ such that they admit a Taylor expansion of the form $\sum_{k=0}^\infty a_kz^k$, where $N^ka_k$ is an integer.
\end{itemize}
The latter is expected  from the relation of periods to the 
integrality of BPS states for smooth Calabi-Yau manifolds without
torsion from mirror symmetry (see, e.g., refs.~\cite{Greene:1993vm,Klemm:2007in}) and observed in 
an overwhelming number of  cases, but, except for a few cases, not proven.

While many Picard-Fuchs operators describing one-parameter families of CY varieties with Hodge structure $(1,1,\ldots,1,1)$ fall into this class, we emphasise that the two concepts are distinct. In particular, for many examples of CY operators of degree four~\cite{Almkvist1,Almkvist2}, it is not known if  they  are Picard-Fuchs operators of some family of CY three-folds.  One-parameter families whose Hodge structure is not $(1,1,\ldots,1,1)$, do not have almost Calabi-Yau operators as Picard-Fuchs operators. For example, the one-parameter families of CY four-folds considered in ref.~\cite{Gerhardus:2016iot}. As we will see, the one-parameter families of Calabi-Yau manifolds $\mathcal{M}_{M,N}$ with $M\le N$, $M\ge 2$  and $N>2$  fulfil neither $(i)$ nor $(ii)$.

\subsection{Operations on differential operators}
\label{eq:opsonops}
In this section, we review standard techniques, such as the Hadamard product, as well 
as the symmetric and anti-symmetric products. They allow one to start 
from a given differential system $L f(z) =0$ and its solutions and to construct new ones. These techniques 
have been studied intensively in the context of differential motives. As we will see in section~\ref{sec:PF_fishnets}, 
these products allow us to relate the periods describing the CY varieties $\mathcal{M}_{M,N}$ for fishnet integrals for different values of $(M,N)$.

In the following, it will be useful to consider 
two differential systems $L f(z) =0$ and $\tilde L f(z) =0$ as equivalent, if the dependent function $f(z)$ is changed by 
an algebraic function $\alpha(z) \in {\overline {\mathbb{Q}(z)}}$.\footnote{If $\alpha(z) \notin {\mathbb{Q}(z)}$, it is referred to
as a twist of the differential motive. It gives only an overall factor to all monodromies, 
hence does not affect irreducibility questions etc.}
In other words, we consider equivalence classes ot differential operators defined as follows: we say that $\tilde{L}$ and ${L}$ are equivalent, $\tilde{L}\sim L$ if 
\beq 
\tilde L = \alpha(z) L \alpha(z)^{-1}, \qquad \alpha(z) \in {\overline {\mathbb{Q}(z)}}\, .
\label{eq:equivalence}
\eeq

We start by defining the {Hadamard product}. Given two functions $f$ and $g$ that are holomorphic at $z=0$,
\beq
f(z) = \sum_{i=0}^\infty f_i z^i\textrm{~~~and~~~}g(z) = \sum_{i=0}^\infty g_i z^i\,,
\eeq
their \emph{Hadamard product} is defined as
\beq
(f\ast g)(z) \coloneqq \oint_{|z|=\epsilon}\frac{\rd t}{2\pi i t}f(t)g(z/t) = \sum_{i=0}^{\infty} f_ig_iz^i\,.
\eeq
Moreover, consider the differential operators $L_f$ and $L_g$ of minimal degree that annihilate $f$ and $g$, respectively.
Their Hadamard product is then defined by
\beq
L_f\ast L_g := L_{f\ast g}\,.
\eeq
Hadamard products have been extensively studied to generate new Calabi-Yau 
operators\cite{MR3822913,Almkvist1} from old ones. They come geometrically with multi fibre structure~\cite{SchoenCY}, whose singularities allow one to predict if the resulting 
geometry is Calabi-Yau and to calculate their Hodge numbers \cite{MR3141723,MR2363121} 
and other topological data. 

Let us now define symmetric and anti-symmetric products of differential operators. This is most conveniently done by considering symmetric and anti-symmetric products of the representation of the monodromy group $\oG$ on $\Sol(L)$. 
If the periods, which are in $\Sol(L)$, form an irreducible fundamental representation  
of $\oG$, 
we call the latter 
$\mV=V^{w_1}$, following our main reference in representation theory~\cite{MR1153249}, where also
the weights $w_i$ are defined. 
The $m^{\textrm{th}}$ symmetric and anti-symmetric power representations will respectively be denoted by $\Sym^m{\mV}$ and $\wedge^m \mV$. As explained below, the latter are represented by spaces of functions constructed from the solutions of $L$, and we identify the representations of $\oG$ with these spaces of functions. The equivalence classes of minimal irreducible operators that annihilate these function spaces are called $\Sym^mL$ and $\wedge^m L$, respectively. 

The solution spaces of $\Sym^mL$ and $\wedge^m L$ can be described very explicitly. Let us start by describing the solution space of the symmetric product. $\Sym^mL$  is defined to be the operator of minimal degree that 
annihilates all products of $m$ solutions of $L$, i.e., it is defined by its 
solution space
\beq
\Sol(\Sym^mL) = \Sym^m \mV=\big\langle y_{i_1}\cdots y_{i_m}: y_{i_k}\in \Sol(L)\big\rangle_{\mathbb{C}}\, .
\eeq
This vector space realises in an obvious sense the symmetric  power of the fundamental representation of $\overline{\mathcal G}$.
If $L$ has degree $n$, then $\dim\Sol L=n$, and we have
\beq
\dim\Sol(\Sym^mL) = \binom{n+m-1}{m}\,.
\eeq

Next, let us describe the solution space of the anti-symmetric (or exterior) power $\wedge^mL$. We start by noting that $z$ is monodromy invariant, 
and so $\mV$ and $\theta_z^k\mV$ are in the same monodromy representation. Hence, the anti-symmetric  
power of the fundamental representation can be represented by the set of 
$\left(n\atop m\right)$ functions represented by $\theta_z^{j_1} \mV \wedge\cdots \wedge
\theta_z^{j_m}\mV$, where the $j_k$, $k=1,\ldots,m$, take ordered, non-repeating values in $\{0,\ldots,n-1\}$. Note that this definition does not depend on the choice of $J=(j_1,\ldots,j_m)$, as all choices lead to the same representation of 
$\oG$, and so the corresponding functions can only differ by $\alpha(z) \in {\mathbb{Q}(z)}$ and we call the antisymmetric 
power  simply $\Lambda^m \mV$.  Let $A_{m,n}$ be the set of $\left(n\atop m\right)$ $m$-tuples $J$ as defined above.
For a given choice, e.g. $J=(0,\ldots,m-1)$, we represent 
the $\left(n\atop m\right)$ functions spanning $\Lambda^m \mV$ as determinants of $m\times m$ minors 
\beq\label{eq:det_form}
D_{I}^J := \det  \left(\begin{smallmatrix} \theta_z^{j_1}y_{i_1} & \ldots & \theta_z^{j_1} y_{i_m} \\ 
\theta_z^{j_2} y_{i_1} &  & \theta_z^{j_2} y_{i_m} \\
\vdots & \ddots & \vdots\\
\theta_z^{j_m} y_{i_1} &\cdots  & \theta_z^{j_m}y_{i_m}
\end{smallmatrix}\right)\,
\eeq
of the $\theta$-Wronskian $W$ with elements $(W)_{ij}=\theta^j_z y_i$, $i,j=0,\ldots, n-1$ and $y_i\in \mV$. We also define $D_I = D_I^{(0,\cdots, m-1)}$.\footnote{Conventionally, one uses 
$\partial_z^k\mV $ and the standard Wronskian instead of $\theta_z^k\mV $ and the  
$\theta$-Wronskian. However, the latter has the advantage that the functions in $\wedge^k \mV$ 
have no poles at the $z=0$ MUM-point of eq.~\eqref{eq:hypergeometricmotive}.}   
The $m^{\textrm{th}}$ anti-symmetric power of $L$ is then defined as the irreducible operator of minimal degree with solution space
\beq
\Sol(\wedge^m L)=\wedge^m \mV = \big\langle D_{I}: I \in A_{m,n}\big\rangle_{\mathbb{C}}\, .
\eeq
Let us make two comments. First, note that we only need 
to consider $m\le \frac{n}{2} $ because  
\beq\label{eq:lambda_relations}
\wedge^{n-m}L \sim \wedge^m L\,.
\eeq
Second, the $D_I$ generate the solution space, i.e., every solution can be written as a linear combination of such determinants. In general, however, the determinants do not necessarily form a basis of $\Sol(\wedge^m L)$, but there may be relations among them. 

If $\mV$ is in an irreducible representation of 
the geometric monodromy group ${\overline{ \cal G}}$ as in \eqref{eq:Hodgegroups}\eqref{G2}, then representation theory decomposes $\wedge^m \mV$ into irreducible representations of ${\overline{\cal G}}$. By identifying one solution in such an irreducible  representation, we immediately find the number of independent solutions corresponding to variations of Hodge structure of the geometry that corresponds to $\wedge^m \mV$ and the representation of the  monodromy action on the associated periods.  We use this fact to analyse the structure of  solutions for the general fishnet integrals in eq.~\eqref{sec:periodsgeneral}.


\section{Period geometry of the  fishnet integrals in two dimensions}
\label{sec:PF_fishnets}
In section \ref{sec:Picardfuchs} we describe how we can compute the Picard-Fuchs operator for the Calabi-Yau varieties given by double coverings like in eq.~\eqref{eq:fishnet_model}, or more specifically  eq.~\eqref{eq:Fishnetgeometry}. Using our general results in section \ref{sec:DEQ_ops}, we explain in section \ref{sec:SymAndHad} how the $\mathcal{M}_{1,N}$
geometries are connected by symmetric and Hadamard products and 
finally in section \ref{sec:periodsgeneral} how the general fishnet period geometry of   $\mathcal{M}_{M,N}$ follows form antisymmetric products of the $\uPi_{1,N}$. 

\subsection{The Picard-Fuchs operators for branched covers of fishnet type}
\label{sec:Picardfuchs}
In order to study the period geometry of $\mathcal{M}_{M,N}$, it is useful to have the expressions for the Picard-Fuchs operator $L_{M,N}$. We now describe our method to determine them in some generality. 

We know that $\Omega_{M,N}$ is the holomorphic $(\ell,0)$-form on $\mathcal{M}_{M,N}$ (with $\ell=MN$). The compact  Calabi-Yau $\ell$-fold \eqref{eq:PiSPi} is given by the double covering of $\bigtimes_{i=1}^\ell \mathbb{P}_i^1$ branched at the  four points in each $\mathbb{P}^1$ at which $\Omega_{M,N}$ can develop a residue. We can make a choice of the branch cuts such that circles $S^1$ in each $\mathbb{P}^1$ lift to non-intersecting circles on the cover to a cycle with topology of an $\ell$-dimensional torus $T^\ell$ on ${\cal M}_{M,N}$.   By taking the residue at say $x_i=0$, $1\le i\le \ell$, it is possible to evaluate an expression for one particular period:
\beq
\label{eq:periodholomorphic}
{\Pi}_{M,N,0}(z) = \oint_{T^\ell}\Omega_{{M,N}}\,.
\eeq
The relation of the  ladder integrals in eq.~\eqref{eq:ladder_kazakov} to the geometries in eq.~\eqref{eq:geomrealisation} can be checked in a stronger 
way at the level of all periods. Since \eqref{eq:geomrealisation} is smooth and the topology is known, we can evaluate eq.~\eqref{eq:PiSPi} using the $\widehat \Gamma$-class and  find eq.~\eqref{eq:ladder_kazakov}, up to the factor $|F(\ua)|^2$.

To compute the special period ${\Pi}_{M,N,0}(z)$, we use a conformal transformation to set $a_1=1,a_2=0,a_4=\infty$ such that we can identify the external point $a_3$ with the cross ratio $z$, i.e.\ $a_3=z$. In this specific configuration of the external points, we can compute ${\Pi}_{M,N,0}(z)$ by a systematic expansion for which we will use multiple times the well-known formula of the Taylor expansion of a square root
\begin{equation}
\frac{1}{\sqrt{1-u}} = \sum_{i=0}^\infty \binom{2i}{i} \left(\frac{u}{4}\right)^i\, ,\quad |u|<1\,.
\label{eq:taylor}
\end{equation}
The integration cycle of ${\Pi}_{M,N,0}(z)$ is just given by an $\ell$-dimensional torus, which means that we have to compute an $\ell$-dimensional residue. This residue can be computed quite easily if we first factor out the product of all the integration variables $x_i$ 
\begin{equation}
 {\Pi}_{M,N,0}(z) = \oint_{T^\ell}\Omega_{{M,N}} \xrightarrow[\text{trans.}]{\text{conf.}} \oint_{T^\ell} \prod_{i=1}^\ell \mathrm dx_i \, \frac{1}{\sqrt{{\hat P}_{M,N}(x,z)}} = \oint_{T^\ell} \prod_{i=1}^\ell \frac{\mathrm dx_i}{x_i}\, \frac{1}{\sqrt{\tilde P_{M,N}(x,z)}} \, ,
\label{eq:conftrans1}
\end{equation}
where the polynomial ${\hat P}_{M,N}(x,z)$ is obtained from $P_{M,N}(x,a)$ in eq.~\eqref{eq:P_G_def} after setting the external parameters to the values $0,1,z,\infty$, and we define $\tilde P_{M,N}(x,z) = \hat P_{M,N}(x,z)\prod_{i=1}^\ell x_i^{-2}$.\footnote{
To compensate for the linear factors that become infinite in the limit $a_4\to\infty$, one has to multiply by a suitable prefactor.} Now we can use eq.~\eqref{eq:taylor} for every linear factor in the polynomial $\tilde P_{M,N}(x,z)$ separately. Then the residue in eq.~\eqref{eq:conftrans1} only receives contributions from the first term containing the product of all $x_i$ variables, if and only if, we pick out in the subsequent expansions the constant term in all $x_i$ variables. In this way we obtain a Taylor series representation for the period close to $z=0$:
\beq\label{eq:4-point-holomorphic-Taylor}
 {\Pi}_{M,N,0}(z)  = \sum_{i=0}^\infty r_{M,N}^{(i)}z^i\,.
 \eeq

Let us present some examples to illustrate this procedure. Our first example is the one-loop ladder integral. In this case we find
\begin{equation}
    \begin{aligned}
        &\oint_{T^1} \mathrm dx \frac1{\sqrt{x(1-x)(x-z)}} = \oint_{T^1} \frac{\mathrm dx}{x} \frac1{\sqrt{(1-x)(1-z/x)}} \\
        &\quad = \oint_{T^1} \frac{\mathrm dx}{x} \sum_{i,j} \binom{2i}{i}\binom{2j}{j}\frac{z^j}{4^{i+j}}x^{i-j} = 2\pi i \sum_{i=0}^\infty \binom{2i}{i}^2 \left(\frac{z}{4^2}\right)^i \, .
    \end{aligned}
\end{equation}
As a second, less trivial example we consider the $(2,2)$-fishnet integral. We start by noticing that
\begin{equation}
    \begin{aligned}
    \tilde P_{2,2}(x,z) &= \big(1-x_{11}\big)\big(1-x_{21}\big)\big(1-\tfrac{z}{x_{12}}\big)\big(1-\tfrac{z}{x_{22}}\big)  \big(1-\tfrac{x_{12}}{x_{11}}\big)\big(1-\tfrac{x_{12}}{x_{22}}\big)\big(1-\tfrac{x_{22}}{x_{21}}\big)\big(1-\tfrac{x_{11}}{x_{21}}\big)\,,
    \end{aligned}
\end{equation}
such that we get
    \begin{align}
        &\oint_{T^4}\frac{\mathrm dx_{11}\mathrm dx_{12}\mathrm dx_{21}\mathrm dx_{22}}{x_{11}x_{12}x_{21}x_{22}}\, \frac1{\sqrt{\tilde P_{2,2}(x,z)}} \\
 \nonumber       &= \oint_{T^4}\frac{\mathrm dx_{11}\mathrm dx_{12}\mathrm dx_{21}\mathrm dx_{22}}{x_{11}x_{12}x_{21}x_{22}}\, \sum_{i_1,\hdots,i_8}\prod_{j=1}^8 \binom{2i_j}{i_j}\left(\frac14\right)^{i_j}z^{i_3+i_4}\,
        x_{11}^{i_1-i_5+i_8}x_{12}^{-i_3+i_5+i_6}x_{21}^{i_2-i_7-i_8}x_{22}^{-i_4-i_6+i_7} \\
 \nonumber       &=(2\pi i)^4\sum_{n=0}^\infty a_n \left(\frac{z}{4^2}\right)^n\,,
        \end{align}
with
    \begin{align} 
    \nonumber    a_n&=\sum_{k_1,k_2,k_3=0}^n \binom{2k_1}{k_1}\binom{2k_2}{k_2}\binom{2k_3}{k_3}
        \binom{2(n-k_1)}{n-k_1}\binom{2(n-k_2)}{n-k_2} \binom{2(n-k_3)}{n-k_3}\binom{2(k_2-k_3)}{k_2-k_3}\\
        &\quad\times\binom{2(k_1+k_3-n)}{k_1+k_3-n} \left(\frac14\right)^{k_1+k_2} \, .
    \end{align}
 
 The Picard-Fuchs operator can now be obtained as follows: Consider the differential operator $L_{M,N}^0$ of minimal degree that annihilates ${\Pi}_{M,N,0}(z)$. We can construct it explicitly by writing down an ansatz for $L_{M,N}^0$ in the form~\eqref{eq:L_standard} and determine the free coefficients in the ansatz so that $L_{M,N}^0$ annihilates the Taylor expansion in eq.~\eqref{eq:4-point-holomorphic-Taylor} up to the order through which we have determined the $r_{M,N}^{(i)}$. We can check that we obtain the same answer if we compute additional Taylor coefficients. Since the Picard-Fuchs operator $L_{M,N}$ annihilates ${\Pi}_{M,N,0}(z) $, it must lie in the ideal generated by $L_{M,N}^0$, i.e., it must be of the form $L_{M,N} = L'L_{M,N}^0$ of some differential operator of degree $s$. If $s>0$, the Picard-Fuchs operator factorises, which implies that the monodromy representation on the periods is reducible. Since CY manifolds are expected to have irreducible monodromy, we conclude that we can pick $L_{M,N} = L_{M,N}^0$. Using this approach, we can construct the Picard-Fuchs operators on a case by case basis. For the ladder integrals, the calculation can be done systematically to find that the series $\Pi_{1,N,0}$ is hypergeometric and $L_{1,N}$ is given in eq.~\eqref{eq:hypergeometricmotive}. By the theorem of Beukers and Heckman,  their geometric monodromy is given in eq.~\eqref{eq:Hodgegroups} acting on its periods. This fixes the variation of Hodge structure of the general $(M,N)$ as antisymmetric powers, as explained in section \ref{sec:periodsgeneral}. 

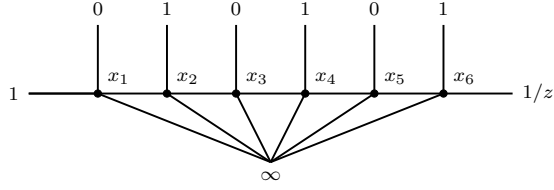
\begin{figure}[!h]
\centering
 \resizebox{0.45\textwidth}{!}{%
\begin{tikzpicture}

\coordinate (p0) at  (0,.5);
\node (p0) at  (p0) [font=\scriptsize, left =0 of p0] {$1$};

\coordinate (p1) at  (1,1.5);
\node (p1) at  (p1) [font=\scriptsize, above =0 of p1] {$0$};

\coordinate (x1) at  (1,.5);
\draw[fill=black] (x1) circle (1.5pt); 
\node (x1) at  (x1) [font=\scriptsize, above right=0 of x1] {$x_1$};

\coordinate (p2) at  (2,1.5);
\node (p1) at  (p2) [font=\scriptsize, above =0 of p2] {$1$};

\coordinate (x2) at  (2,.5);
\draw[fill=black] (x2) circle (1.5pt); 
\node (x2) at  (x2) [font=\scriptsize, above right=0 of x2] {$x_2$};

\coordinate (p3) at  (3,1.5);
\node (p3) at  (p3) [font=\scriptsize, above =0 of p3] {$0$};

\coordinate (x3) at  (3,.5);
\draw[fill=black] (x3) circle (1.5pt); 
\node (x3) at  (x3) [font=\scriptsize, above right=0 of x3] {$x_3$};

\coordinate (p4) at  (4,1.5);
\node (p4) at  (p4) [font=\scriptsize, above=0 of p4] {$1$};

\coordinate (x4) at  (4,.5);
\draw[fill=black] (x4) circle (1.5pt); 
\node (x4) at  (x4) [font=\scriptsize, above right=0 of x4] {$x_4$};

\coordinate (p5) at  (5,1.5);
\node (p5) at  (p5) [font=\scriptsize, above=0 of p5] {$0$};

\coordinate (x5) at  (5,.5);
\draw[fill=black] (x5) circle (1.5pt); 
\node (x5) at  (x5) [font=\scriptsize, above right=0 of x5] {$x_5$};

\coordinate (p6) at  (6,1.5);
\node (p6) at  (p6) [font=\scriptsize, above=0 of p6] {$1$};

\coordinate (x6) at  (6,.5);
\draw[fill=black] (x6) circle (1.5pt); 
\node (x6) at  (x6) [font=\scriptsize,  above right=0 of x6] {$x_6$};

\coordinate (iz) at  (7,.5);
\node (iz) at  (iz) [font=\scriptsize, right=0 of iz] {$1/z$};

\coordinate (inf) at  (3.5,-.5);
\node (inf) at  (inf) [font=\scriptsize, below =0 of inf] {$\infty$};

 \draw[-,black!100,thick] (3.5,-.5) to (1,.5);
\draw[-,black!100,thick] (3.5,-.5) to (2,.5);
\draw[-,black!100,thick] (3.5,-.5) to (3,.5);
\draw[-,black!100,thick] (3.5,-.5) to (4,.5);
\draw[-,black!100,thick] (3.5,-.5) to (5,.5);
\draw[-,black!100,thick] (3.5,-.5) to (6,.5);
\draw[-,black!100,thick] (p0) to (1,.5);
\draw[-,black!100,thick] (1,1.5) to (1,.5);
\draw[-,black!100,thick] (p2) to (2,.5);
\draw[-,black!100,thick] (p3) to (3,.5);
\draw[-,black!100,thick] (p4) to (4,.5);
\draw[-,black!100,thick] (p5) to (5,.5);
\draw[-,black!100,thick] (p6) to (6,.5);
\draw[-,black!100,thick] (0,.5) to (7,.5);
\end{tikzpicture}}
\caption{Non-planar six-loop  non-oriented fishnet  graph leading to
an irreducible mondromy representation in the Lie group $\overline {\cal G}=G_2$.}
\label{fignonoriented}
\end{figure}

 Let us give finally an example of a non-planar fishnet graph, 
shown in figure \ref{fignonoriented}. This graph gives rise  to a one-parameter variation of a Calabi-Yau six-fold of type $(1,\ldots,1)$ that has  been very well studied and is famous in mathematics since the Zariski closure of its geometric monodromy group $ {\cal G}$ is in the Lie group $G_2$ \cite{MR2660679,MR3484368}
 \beq\label{G2}
 {\overline {\cal G}}=G_2 \ .
 \eeq
It can also be realized as a singular double covering of 
 $\bigtimes_{i=1}^6 \mathbb{P}_i^1$ and the analogous equation to eq.~\eqref{eq:Fishnetgeometry} is now given by \cite{MR3484368}
 \beq\bsp\label{eq:G2}
y^2&=x_1(1-x_1)(x_1-x_2)(1-x_2)(x_2-x_3)x_3(x_3-x_4)(1-x_4)\\ &(x_4-x_5)x_5 (x_5-x_6)(1-x_6)(1-z x_6)\, ,
\esp\eeq
which is readily identified with the non-planar fishnet graph depicted 
in figure \ref{fignonoriented}.
Performing the residue integrals in eq.~\eqref{eq:periodholomorphic} 
we get a particular combination of binomials from which we infer 
that the Picard Fuchs operator reads 
\begin{equation}
\begin{aligned}
\mathcal L_7 =   &\theta_z^7\!- \!\frac{z}{2^7} (2 \theta_z \!+\!1) (256\theta_z^6+768 \theta_z^5+1312 \theta_z^4+1344\theta_z^3+832 \theta_z^2+288 \theta_z+43)  \\  
                & +\frac{3 z^2}{2^7}\!\!\prod_{k=1}^3 (2 \theta_z\!+\!k)(32\theta_z^4+128\theta_z^3+240\theta_z^2+224\theta_z+87) \\
                & -\frac{z^3}{2^6}\prod_{k=1}^5 (2 \theta_z\!+\!k)(8\theta_z^2+24\theta_z+25) +\frac{z^4}{2^7}\prod_{k=1}^7 (2 \theta_z\!+\!k)\,.
\label{eq:L7}
\end{aligned}
\end{equation}
This operator has not appeared in the literature before. Note, however, that the 
Riemann symbol of $ \mathcal L_7$ given is by
\begin{equation}
\begin{aligned}
	\mathcal P	&	\begin{Bmatrix*}[l]	%
							z 				& 1 & \infty			\\ \hline
							0 &0 &\frac{1}{2}	\\
                            0 &0 &\frac{3}{2}	\\
                           0 &1 &\frac{5}{2}	\\
                           0 &1 &\frac{7}{2}	\\
                           0 &1 &1	\\
                             0 &2 &2	\\
                             0 &2 &3	\\ 
        \end{Bmatrix*} \,, 
\label{Riemannepseven}
\end{aligned}
\end{equation}
and its local monodromy agrees with properties of the monodromy blocks that 
have been calculated in ref.~\cite{MR2660679} by analysing the local 
action on the homology of eq.~\eqref{eq:G2} induced by monodromy paths in the $z$ plane around the singular fibres at $z=0,1,\infty$.

 
\subsection{Period geometries from symmetric and Hadamard products} 
\label{sec:SymAndHad}
We now discuss relations between the Picard-Fuchs operators $L_N=L_{1,N}$  for different values of $N$. It is easy to see that we have
\beq
{\Pi}_{N,0} = {\Pi}_{p,0} \ast {\Pi}_{N-p-1,0}\,, \quad 0\le p \le N-1\,,
\eeq
where we defined
\beq
{\Pi}_{0,0}(z) = \sum_{i=0}^\infty \binom{2i}{i}\,\left(\frac{z}{4}\right)^i = \frac{1}{\sqrt{1-z}}\,.
\eeq
From the discussion in the previous section, we see that the Picard-Fuchs operators for ladder integrals can be constructed iteratively as Hadamard products. In particular, adding one more integration vertex corresponds to taking the Hadamard product with $L_0$:
\beq\label{eq:L_Hadamard}
L_N = L_p\ast L_{N-p-1} = L_0\ast L_{N-1} = L_0^{\ast (N+1)}\,, \quad 0\le p \le N-1\,.
\eeq

 It is known that every CY operator of degree 3 is equivalent to the symmetric square of a CY operator of degree 2~\cite{doran,BognerCY,BognerThesis}. We find that
\beq
L_2=\Sym^2(L_1)\,.
\eeq
Since the solutions of CY operators of degree 2 are periods of a family of elliptic curves, we conclude that the periods of $\mathcal{M}_{1,1}$ and $\mathcal{M}_{1,2}$ can be expressed in terms of complete elliptic integrals of the first kind.
The one-loop ladder integral was discussed in refs.~\cite{Derkachov:2018rot,Corcoran:2021gda} in terms of elliptic integrals. Indeed, we find
\beq
\Sol(L_1) = \big\langle \Pi_{1,0}(z), \Pi_{1,1}(z)\big\rangle_{\mathbb{C}}\,,
\eeq
with
\beq\label{eq:PiTilde_G11}
\Pi_{1,0}(z) =  \frac{2}{\pi}\,\K(z)\,,\quad  
\Pi_{1,1}(z) =  \frac{2}{\pi}\,\K(1-z)\,,
\eeq
and $\K(z)$ denotes the complete elliptic integral of the first kind:
\beq
\K(z) = \int_0^1\frac{dt}{\sqrt{(1-t^2)(1-zt^2)}}\,.
\eeq
 For $\mathcal{M}_{{1,2}}$, we find~\cite{Duhr:2022pch}:
\beq
\Sol(L_2) = \big\langle \Pi_{2,0}(z), \Pi_{2,1}(z), \Pi_{2,2}(z)\big\rangle_{\mathbb{C}}\,,
\eeq
with
\beq
\label{sol2}
\Pi_{2,0}(z) =  \frac{4}{\pi^2}\,\K(z_-)^2\,,\quad  
\Pi_{2,1}(z) =  \frac{4}{\pi^2}\,\K(z_-)\,\K(z_+)\,, \quad
\Pi_{2,2}(z) =  \frac{4}{\pi^2}\,\K(z_+)^2\,,
\eeq
and
\beq
z_{\pm} = \frac{1}{2}(1\pm \sqrt{1-z})\,.
\eeq
For $N>2$ it is not expected that the periods can be expressed in terms of elliptic integrals. However, by mirror symmetry we have interesting
integral structures in geometric $q$ expansions of the periods (and the 
quantum volume in eq.~\eqref{eq:Vol_q_expanded}) at the points of maximal unipotent monodromy. We discuss the  higher dimensional geometries
below.


\subsection{The period geometry for general fishnet integrals}
\label{sec:periodsgeneral}
Let us now turn to the genuine fishnet integrals with $M>1$. We have computed the periods in eq.~\eqref{eq:4-point-holomorphic-Taylor} up to $(M,N)=(2,4)$ and $(3,3)$. We find that in all cases we can write (cf.~eq.~\eqref{eq:det_form}):
\beq
\label{detDI}
\Pi_{M,N,0}(z) = {D}^{({W})}_{01\cdots (M-1)}(z)\,,
\eeq
where we defined ${W}:=M+N-1$ and 
\beq\label{eq:det_periods}
{D}^{({W})}_{I}(z) := \det\left(\begin{smallmatrix} \phantom{\theta_z }\Pi_{{W},i_1}(z) & \ldots & \phantom{\theta_z }\Pi_{{W},i_M}(z) \\ 
\theta_z \Pi_{{W},i_1}(z) &  & \theta_z\Pi_{{W},i_M}(z) \\
\vdots & \ddots & \vdots\\
\theta_z^{M-1} \Pi_{{W},i_1}(z) &\cdots  & \theta_z^{M-1}\Pi_{{W},i_M}(z)
\end{smallmatrix}\right)\,,\qquad I = (i_1,\ldots,i_M)\,.
\eeq 
From this we see that the Picard-Fuchs operator of $\mathcal{M}_{{M,N}}$ is the $M^{\textrm{th}}$ exterior power of $L_{W}$,
\beq\label{eq:det_fishnet}
L_{M,N} \sim \wedge^{M}L_{{W}} \sim \wedge^ML_0^{\ast({W}+1)}\,,
\eeq
where the last equivalence follows from eq.~\eqref{eq:L_Hadamard}. Note that this relation remains true for $M=1$. We see that the Picard-Fuchs operators for $\mathcal{M}_{{M,N}}$ can be constructed by taking Hadamard and exterior products of the first-order operator $L_0$.

Equation~\eqref{eq:det_fishnet} has interesting implications for fishnet graphs. First, eq.~\eqref{eq:lambda_relations} implies that $L_{M,N}$ and $L_{N,M}$ are equivalent:
\beq\label{eq:L_MN_symmetry}
L_{M,N} \sim \wedge^{M}L_{{W}} \sim \wedge^{{W}-M+1}L_{{W}} = \wedge^{N}L_{{W}} \sim L_{N,M}\,.
\eeq
This implies that $\mathcal{M}_{{M,N}}$ and $\mathcal{M}_{{N,M}}$ have the same periods, up to multiplication by an algebraic function of $z$. This is not unexpected from the Feynman diagram perspective, because the corresponding Feynman diagrams are simply related by a permutation of the external points. Another consequence is that we can limit ourselves to the cases $M\le N$.

Equation~\eqref{detDI}, or more generally eq.~\eqref{eq:det_fishnet} subject to eq.~\eqref{eq:L_MN_symmetry}, implies
\beq
 L_{M,N} \sim \wedge^M L_{1,M+N-1}\ ,
\eeq
where without restriction of generality $M\le N$.  We can now use the results on the geometric monodromy group of the ladder geometries in eq.~\eqref{eq:Hodgegroups} and  the general considerations in sections \ref{eq:opsonops} to characterize  the  solution space $\mV_{M,N}$ of $L_{M,N}$ completely within the equivalence defined by  \eqref{eq:equivalence} from the solution space $\mV$ of  $L_{1,M+M}$ using representation theory. The choice of the  $\alpha(z)\in 
{\overline {\mathbb{Q} (z)}}$ must still be made based on physically 
preferred representatives.  
Let us start with the dimension of the solution space $\Sol(\wedge^M L_{1,W})$. For $W=M+N-1$ even and $M<\frac{W}{2}$ (or $M=\frac{W}{2}$), it follows from 
Theorem 19.4 of ref.~\cite{MR1153249}\footnote{For notations for the  weight vectors $w_k$ we follow the convention of~\cite{MR1153249}.} that $\wedge^M \mV=V^{w_M}$ (or  $\wedge^M \mV=V^{2 w_M}$). In particular, this representation is irreducible and of rank $r=\left(M+N\atop M\right)$. For $W=M+N-1$ odd, it follows from Theorem 17.5 of ref.~\cite{MR1153249} that the largest irreducible sub-representation of $\wedge^M \mV$  is $V^{w_M}=\wedge^M \mV/\wedge^{M-2} \mV$, with rank $r=\left(M+N\atop M\right)-\left(M+N\atop M-2\right)$. We can summarise this by:
\beq\label{eq:dimSolLMN}
\dim\Sol(L_{M,N})  = \deg L_{M,N} = \left\{\begin{array}{ll}
\binom{M+N}{M}\,,& \textrm{ if } M+N \textrm{ odd}\,,\\
\phantom{a}\\
\binom{M+N}{M}-\binom{M+N}{M-2}\,,& \textrm{ if } M+N \textrm{ even}\,.
\end{array}\right.
\eeq
From this we deduce that for $M+N$ odd, the determinants $D_I^{(W)}$ are linearly independent, while for $M+N$ even there must be $\binom{M+N}{M-2}$ relations among them. Indeed, we have already stated that, while the determinants in eq.~\eqref{eq:det_periods} generate $\Sol(L_{M,N})$, they do not necessarily form a basis for it. Thanks to eq.~\eqref{eq:L_MN_symmetry}, it is sufficient to discuss the case $M\le \frac{W-1}{2}$. In appendix~\ref{app:proof_rels} we show that, as a consequence of the anti-symmetry of the intersection form for $M+N$ even and the quadratic relations among periods due to eq.~\eqref{Griffiths}. We have the following relations for $W = M+N-1$  odd and $M\le \frac{W-1}{2}$:
\beq\label{eq:det_relations}
\sum_{k=0}^{\frac{W-1}{2}}(-1)^k\,D^{(W)}_{k(W-k)J}(z) = 0\,,\textrm{~~for all } J\in T_{W,M-2} = \{(i_1,\ldots, i_{M-2}) : 1\le i_k\le W\}\,.
\eeq
For example, for $(M,N)=(2,2)$, there are 6 determinants, but we have (with ${W}=3$)
\beq\label{eq:det_4_rel}
{D}^{(3)}_{03}(z) = {D}^{(3)}_{12}(z)\,,
\eeq
and so only five out of six determinants are linearly independent~\cite{Almkvist3}. Note that there are precisely $\binom{M+N}{M-2}$ relations of the form~\eqref{eq:det_relations} one can write down. Typically, for a given choice of $J$, many terms in eq.~\eqref{eq:det_relations} vanish, so one may wonder if some of these relations may be trivially satisfied, e.g., because all determinants in the sum vanish individually. It is easy to see that this will never be the case. Indeed, the determinants would vanish individually if for example $k\in J$, for all $0\le k\le \frac{W-1}{2}$. This is impossible, because $J$ contains $M\le \frac{W-1}{2}$ elements, but the sum in eq.~\eqref{eq:det_relations} contains $\frac{W+1}{2}$ terms. Hence, there is always at least one term in eq.~\eqref{eq:det_relations} that is non-zero, and we obtain $\binom{M+N}{M-2}$ relations.

Note that by definition $\dim\Sol(L_{M,N})$ equals the rank $b_{M\cdot N}$ of the middle cohomology of $\mathcal{M}_{M,N}$ (cf.~eq.~\eqref{eq:period_vector}). 
By comparing the Hodge filtration as well as the Tate filtration of 
$\mathcal{M}_{1,W}$ with that of $\mathcal{M}_{M,N}$, and using eq.~\eqref{eq:det_fishnet} and the  
irreduciblility of 
$\oG_W$ for all $L_{W}$, we can refine eq.~\eqref{eq:dimSolLMN} to give 
the individual Hodge numbers~\cite{privcomkerr}. We present here a more pedestrian argument, using 
the fact that $L_{W}$ has a MUM-point at $z=0$, with the solution 
structure given in eq.~\eqref{eq:Frobenius} with $\lambda=0$ and $m=l$. Given the fact that 
eq.~\eqref{eq:Frobenius} depends only on the $W+1$ independent power series $\varpi_i(z)$, $i=0,\ldots,W$, 
it is straightforward to count the number of independent functions in $\wedge^M \mV$ (with $\mV=\Sol(L_{W})$) with leading behaviour $\log^kz$ as $z\to0$. These are precisely the Hodge numbers $h^{\ell-k,k}(\mathcal{M}_{M,N})$, 
$k=0,\ldots, b_{\ell}-1$. For $W$ even the number of such solutions is the number $\pi_{M,N}(k)$ 
of partitions of $k$ in sums of $N$ not necessarily distinct integers in the range $[0,M]$.
For $W$ odd, $\pi_{M+2,N-2}(k-2-(M-N))$ of these are dependent, hence 
\beq
\label{eq:hodgeMN}
h^{\ell-k,k}(\mathcal{M}_{M,N})=\left\{\begin{array}{ll}
\pi_{M,N}(k) \,,& \textrm{ if } M+N \textrm{ odd}\,,\\
\phantom{a}\\
\pi_{M,N}(k)-\pi_{M+2,N-2}(k-2-(M-N))\,,& \textrm{ if } M+N \textrm{ even}\,.
\end{array}\right.
\eeq
One can also  give a generating series for the $\pi_{M,N}(k)$:
\beq 
\sum_{k=1}^{MN}\pi_{M,N}(k) T^k=\prod_{j=1}^M \frac{1-T^{M+j}}{1-T^j} \ .
\eeq 
Note that we have (see also appendix~\ref{app:proof})
\beq\label{eq:h_structure}
h^{\ell-k,k}(\mathcal{M}_{M,N}) = 1 \textrm{~~for~~} k=0,1,\ell-1,\ell\,.
\eeq
This is precisely the structure expected from the Hodge numbers of a one-parameter family of CY $\ell$-folds.
For $2\le k\le \ell-2$, we generically have $h^{\ell-k,k}(\mathcal{M}_{M,N}) > 1$, and so $L_{M,N}$ is not a CY operator for $M>1$. 
The values of eqs.~\eqref{eq:dimSolLMN} and \eqref{eq:hodgeMN} for $M+N\le 8$ are tabulated 
in Table \ref{tab:h_p}.
We see that for $(M,N)=(2,2)$, we have $h^{\ell-k,k}(\mathcal{M}_{2,2})=1$ for all $0\le k\le \ell$. We have already seen that the second exterior power of a CY operator of degree 4 is conjectured to be a CY operator of degree 5. Hence, $L_{2,2} = \wedge^2L_3$ is expected to be
a CY operator (cf.,~e.g.,~refs.~\cite{BognerThesis,Almkvist3}). If either $M$ or $N$ is greater than 2, we do not expect that $L_{M,N}$ is a CY operator.

\begin{table}[!t]
\begin{center}
\begin{tabular}{c|c|c|c|c}
\hline\hline
$(M,N)$ & $\ell$& $\binom{M+N}{M}$ & $b_{\ell}(\mathcal{M}_{M,N})$ & $h^{\ell-k,k}(\mathcal{M}_{M,N})$ \\
\hline
$(2,2)$ &4 & 6 & 5 & 1\,\,\,1\,\,\,1\,\,\,1\,\,\,1\\
\hline
$(2,3)$&6 & 10 & 10 & 1\,\,\,1\,\,\,2\,\,\,2\,\,\,2\,\,\,1\,\,\,1\\
\hline
$(2,4)$&8 & 15 & 14 & 1\,\,\,1\,\,\,2\,\,\,2\,\,\,2\,\,\,2\,\,\,2\,\,\,1\,\,\,1\\
$(3,3)$&9 &20 &14&1\,\,\,1\,\,\,1\,\,\,2\,\,\,2\,\,\,2\,\,\,2\,\,\,1\,\,\,1\,\,\,1\\
\hline
$(2,5)$&10 &21 &21&1\,\,\,1\,\,\,2\,\,\,2\,\,\,3\,\,\,3\,\,\,3\,\,\,2\,\,\,2\,\,\,1\,\,\,1\\
$(3,4)$&12 &35 &35&1\,\,\,1\,\,\,2\,\,\,3\,\,\,4\,\,\,4\,\,\,5\,\,\,4\,\,\,4\,\,\,3\,\,\,2\,\,\,1\,\,\,1\\
\hline
$(2,6)$&12 &28 &27&1\,\,\,1\,\,\,2\,\,\,2\,\,\,3\,\,\,3\,\,\,3\,\,\,3\,\,\,3\,\,\,2\,\,\,2\,\,\,1\,\,\,1\\
$(3,5)$&15 &56 &48&1\,\,\,1\,\,\,2\,\,\,3\,\,\,3\,\,\,4\,\,\,5\,\,\,5\,\,\,5\,\,\,5\,\,\,4\,\,\,3\,\,\,3\,\,\,2\,\,\,1\,\,\,1\\
$(4,4)$&16 &70 &42&1\,\,\,1\,\,\,1\,\,\,2\,\,\,3\,\,\,3\,\,\,4\,\,\,4\,\,\,4\,\,\,4\,\,\,4\,\,\,3\,\,\,3\,\,\,2\,\,\,1\,\,\,1\,\,\,1\\
\hline\hline
\end{tabular}
\caption{\label{tab:h_p} The dimension $\ell=MN$ of $\mathcal{M}_{M,N}$, the total number of 
functions in $\wedge^M L_{M+N-1}$, the Betti number $b_{\ell}(\mathcal{M}_{M,N})=\deg L_{M,N} = \dim \Sol(L_{M,N})$, as well its splitting in Hodge numbers $h^{\ell-k,k}(\mathcal{M}_{M,N})$, which 
correspond to the number of solutions with leading logarithmic divergence $\log^kz$ 
at $z=0$ for $M+N\le 8$. }
\label{tab1}
\end{center}
\end{table}

The operators $L_{M,N}$ are polynomials in $z$ and $\theta_z$. 
The degree in $\theta_z$ is $\dim\Sol(L_{M,N})$, and so it is fixed by eq.~\eqref{eq:dimSolLMN}.   
However, we only have estimates for the degree in $z$, and this degree also grows fast. 
For example, the operator     
\beq
\label{L22}
L_{2,2}=\theta_z^5 + z^2 (1 + \theta_z)^5 - \frac{1}{8} z (1 + 2 \theta_z) (3 + 12 \theta_z + 20 \theta_z^2 + 
    16 \theta_z^3 + 8 \theta_z^4)\ ,
\eeq
is the only operator for $M>1$ which is an almost Calabi-Yau operator, and it is quadratic in $z$, and its discriminant is related to the discriminant of $L_{1,3}$,  $\Delta_{L_{2,2}}(z)=\Delta_{L_{1,3}}(z)^2$ and we checked that it fulfils  also  the integer properties $(iii)$. On the other hand, the operators $L_{2,3}=\theta_z^7(\theta_z-1)^3+ {\cal O}(z)$ and $L_{3,3}=\theta_z^{10}(\theta_z-1)^4+ {\cal O}(z)$ are not 
almost Calabi-Yau operators. They are not essentially self-adjoint and 
have no MUM point, but they have the expected local exponents at $z=0$ as given in Table \ref{tab1}. They have maximal order $10$ and $18$ in $z$. From our derivation of eq.~\eqref{eq:hodgeMN}, we can read off the 
degeneracies of all local exponents of $L_{M,N}$ at $z=0$, which by construction obey  
$\lambda_i-\lambda_j \in \mathbb{Z}$. The structure of solutions is as described after eq.~\eqref{eq:Frobenius}.

Since ${\rm SO}(3,2) \twoheadleftarrow {\rm Spin}(3,2)\simeq {\rm SL}(4,\mathbb{R})$ one can show that every exterior power of a CY operator of degree 4 is an almost CY operator of degree 5 with Hodge numbers $(1,1,1,1,1)$~\cite{privcomkerr} and vice versa.
As a consequence, the periods of $\mathcal{M}_{{1,4}}$ are found to be $2\times2$ determinants of periods of some one-parameter family of CY three-folds. We find $L_4 \sim \wedge^2\widetilde{L}_3$, 
with 
\beq\bsp\label{eq:L3_tilde}
\widetilde{L}_3 =& \theta_z^4-2^{4} z(4\theta_z+1)(4\theta_z+3)(32\theta_z^2+32\theta_z+13)\\
&\,+2^{16} z^{2}(4\theta_z+1)(4\theta_z+3)(4\theta_z+5)(4\theta_z+7)\,.
\esp\eeq
This operator corresponds to the entry $2.33$ in the AESZ database of CY operators of degree four~\cite{aesz}. This implies that the periods of $\mathcal{M}_{1,4}$ are $2\times2$ determinants of the solutions of $\widetilde{L}_3$ (and their derivatives). In particular, the solution of $\widetilde{L}_3$ that is holomorphic at $z=0$ reads:
\beq
P_0(z) =  1 + 624\,z+ 1251600\,z^2+ 3268151040\,z^3+ 9627237219600\,z^4+\mathcal{O}(z^5)\,.
\eeq
We have seen that $L_4$ is the second exterior power of the CY operator $\widetilde{L}_3$ in eq.~\eqref{eq:L3_tilde}. In refs.~\cite{Almkvist3,BognerCY,BognerThesis} it was shown that for every CY operator $L$ of degree 4 one has $\wedge^2\wedge^2L \sim \Sym^2L$. We can combine this with eq.~\eqref{eq:det_fishnet} to obtain
\beq
L_{2,3} \sim \wedge^2L_4 \sim \wedge^2\!\wedge^2\widetilde L_3 \sim \Sym^2\widetilde L_3\,.
\eeq
In other words, the periods of the family $\mathcal{M}_{2,3}$ of CY six-folds can be expressed in terms of the periods of the same CY three-fold that appeared in the computation of the periods of $\mathcal{M}_{1,4}$.

\section{The Basso-Dixon formula}
\label{sec:BD}

\subsection{The intersection form for fishnet integrals}

The Picard-Fuchs operators allow us to compute the periods of $\mathcal{M}_{M,N}$. In fact, it is sufficient to compute the periods of the ladder integrals, because all other cases can be obtained by computing determinants of the periods of ladder integrals. The latter are easy to obtain, because the Picard-Fuchs operators for ladder integrals are hypergeometric differential operators. In order to apply eq.~\eqref{eq:PiSPi} to compute the fishnet integrals, we also need the expression for the intersection form $\Sigma_{M,N}$ for $\mathcal{M}_{M,N}$. 

Since the Picard-Fuchs operators for ladder integrals are CY operators, the intersection form for the Frobenius basis is fixed by eq.~\eqref{eq:Sigmal}. 
For $M>1$, the Picard-Fuchs operators are not CY operators (except for $(M,N)=(2,2)$), and a priori it may be a difficult task to write down the intersection form. In the following, we show how one can leverage the knowledge that $L_{M,N} = \wedge^ML_W$ to relate the intersection forms $\Sigma_{M,N}$ and $\Sigma_{W}$.

We start by reviewing some general facts from linear algebra on exterior products. Consider a vector space $V$ of dimension $d+1$ with basis $\{{\bf e}_i: 0\le i\le d\}$. A basis for $\wedge^MV$ is given by the vectors $\widetilde{\bf e}_I:={\bf e}_{i_1}\wedge\ldots\wedge {\bf e}_{i_M}$, 
where $I\in\overline{T}_{d,M} = \{(i_1,\ldots,i_M)\in T_{d,M}: i_{k}<i_{k+1}\}$. Consider an endomorphism $\mathbb{M}\in\End(V)$ represented in this basis by the matrix with entries $M_{ij}$. Then this determines an endomorphism $\widetilde{\mathbb{M}}\in\End\big(\wedge^MV\big)$ represented by the matrix $\widetilde{M}$ with entries
\beq\label{eq:det_endo}
\widetilde{M}_{IJ} = \det M(I,J)\,,
\eeq
where $M(I,J) = \big(M_{ij}\big)_{{i\in I, j\in J}}$ is the matrix obtained from $M$ by only keeping the rows from $I$ and the columns from $J$. It is easy to check that, if $M=M^{(1)}M^{(2)}$, then we have
\beq
\widetilde{M}_{IJ} = \widetilde{M}^{(1)}_{IK}\,\widetilde{M}^{(2)}_{KJ} = \det \left[M^{(1)}(I,K)M^{(2)}(K,J)\right]\,,
\eeq

Consider a set of $d'$ vectors in $V$:
\beq
{\bf v}^{(i)} =c_I^{(i)}\,\widetilde{\bf e}_I\,,\qquad 1\le i\le d'\,,\quad c_I^{(i)}\in\mathbb{C}\,.
\eeq
We assume them to be linearly independent, which means the matrix formed by the coefficients $c_I^{(i)}$ has full rank $d'$. We denote the $d'$-dimensional vector space they span by 
\beq
Q:=\langle {\bf v}^{(i)}:1\le i\le d'\rangle_{\mathbb{C}}\subseteq \wedge^MV\,.
\eeq
Then the endomorphism $\mathbb{M}\in\End(V)$ determines an endomorphism of the quotient $\wedge^MV/Q$ in the following way. Fix a basis $\{\widehat{\bf b}_\alpha\}$ of $\wedge^MV/Q$, e.g., by choosing a maximal linearly independent subset of $\{\widetilde{\bf e}_I\}$.\footnote{In the following, lower-case Greek letters denote indices in $\wedge^MV/Q$, while upper-case Latin letters denote indices in $\wedge^MV$.} Then there is a $d\times(d-d')$ matrix $\widetilde{B}_{I\alpha}$ such that $\widetilde{\bf e}_I = \widetilde{B}_{I\alpha}\widehat{\bf b}_\alpha$. Then $\mathbb{M}$ defines the endomorphism of $\wedge^MV/Q$ described in this basis by the $(d-d')\times(d-d')$ matrix 
\beq\label{eq:M_Hat}
\widehat{M}_{\alpha\beta} = \left[\widetilde{B}^\dagger\,\widetilde{M}\,\widetilde{B}\right]_{\alpha\beta} = \widetilde{B}^*_{I\alpha}\,\widetilde{M}_{IJ}\,\widetilde{B}_{J\beta}\,.
\eeq
At this point we make an important observation. We can use $\widetilde{M}$ and $\widehat{M}$ to define bilinears on $\wedge^MV$ and $\wedge^MV/Q$, respectively, by contracting from the left and right with the appropriate basis vectors. It turns out that the two bilinears agree:
\beq\label{eq:bilinear}
\widehat{\bf b}^*_\alpha\widehat{M}_{\alpha\beta}\widehat{\bf b}_\beta=\left(\widetilde{B}_{I\alpha}\widehat{\bf b}_\alpha\right)^*\,\widetilde{M}_{IJ}\,\left(\widetilde{B}_{J\beta}\widehat{\bf b}_\beta\right) = \widetilde{\bf e}^*_I\widetilde{M}_{IJ}\widetilde{\bf e}_J\,.
\eeq

We can immediately apply these results to construct the intersection form for $\mathcal{M}_{M,N}$ and the bilinear expression in eq.~\eqref{eq:PiSPi}. Indeed, we simply take $V = \Sol(L_{W})$, ${\bf e}_i = \Pi_{W,i}(z)$ and $\widetilde{\bf e}_I = D_I^{(W)}(z)$, 
$I\in\overline{T}_{W,M}$. 
The intersection form $\Sigma_W$ defines an endomorphism on $V=\Sol(L_{W})$. If $M+N$ is odd, then the determinants $D_I^{(W)}(z)$ are linearly independent and form a basis of $V$, and we can construct the intersection form using eq.~\eqref{eq:det_endo}. The case $M+N$ even is slightly more complicated, because the determinants $D_I^{(W)}(z)$ are not linearly independent, and we need to quotient by the relations in eq.~\eqref{eq:det_relations}. The vectors ${\bf v}^{(i)}$ are the linear combinations of determinants in the left-hand side of eq.~\eqref{eq:det_relations}. For fixed $(M,N)$, we can solve these relations and choose a basis $\{\widehat{\bf b}_\alpha\}$ of $\Sol(L_{M,N})\simeq \wedge^MV/Q$, and the intersection form relative to that basis is given by eq.~\eqref{eq:M_Hat}. However, since we are only interested in the bilinear combination in eq.~\eqref{eq:PiSPi}, we can use eq.~\eqref{eq:bilinear} to express this bilinear in terms of the overcomplete set of all determinants $D_I^{(W)}(z)$. As a conclusion, we can write down the following formula for the fishnet integrals, valid for all values of $M+N$:
\beq\label{eq:Phi_MN_Sigma}
\phi_{M,N}(z) = (-i)^{(MN)^2}\,D_I^{(W)}(\bar{z})\,\big(\widetilde{\Sigma}_{M,N}\big)_{IJ}\,D_J^{(W)}({z})\,,
\eeq
where we defined
\beq \label{eq:Sigma_M_N_def}
\big(\widetilde\Sigma_{M,N}\big)_{IJ} = \det\Big[\Sigma_{{W}}(I,J)\Big] = \det\Big[\big(\Sigma_{{W}}\big)_{ij}\Big]_{\substack{i\in I\\ j\in J}}\,.
\eeq
Equation~\eqref{eq:Phi_MN_Sigma} looks very similar to eq.~\eqref{eq:PiSPi}. Its interpretation, however, is different. For $M+N$ odd, the determinants $D_I^{(W)}({z})$ can be identified with a basis $\uPi_{M,N}(z)$ of $\Sol(L_{M,N})$, and $\widetilde\Sigma_{M,N}$ agrees with the intersection form $\Sigma_{M,N}$ on $\mathcal{M}_{M,N}$. If $M+N$ is even, the determinants $D_I^{(W)}({z})$ are not linearly independent, and so they cannot be identified with a basis $\uPi_{M,N}(z)$ of $\Sol(L_{M,N})$. Consequently, $\widetilde\Sigma_{M,N}$ cannot be identified with the intersection form $\Sigma_{M,N}$ on $\mathcal{M}_{M,N}$. Nevertheless, eqs.~\eqref{eq:Phi_MN_Sigma} and~\eqref{eq:PiSPi} agree in all cases, thanks to eq.~\eqref{eq:bilinear}. In particular, we can explicitly check that the bilinear expression in eq.~\eqref{eq:Phi_MN_Sigma} is monodromy-invariant for all values of $M+N$. Let $\rho:\overline{\cG}_{W}\to \End(V)$ be the monodromy representation on $\Sol(L_{W})$. It acts on the determinants $D_I^{(W)}(z)$ via the representation
\beq
\widetilde{\rho}_{IJ} = \det\big[\rho(I,J)\big]\,.
\eeq
Under a monodromy transformation, the bilinear expression in eq.~\eqref{eq:Phi_MN_Sigma} transforms into:
\beq\bsp
D_K^{(W)}(\bar{z})&\,\widetilde{\rho}_{IK}^*\,\big(\widetilde{\Sigma}_{M,N}\big)_{IJ}\,\widetilde{\rho}_{JL}\,D_L^{(W)}({z})\\
&\,=D_K^{(W)}(\bar{z})\,\det\big[{\rho}(I,K)\big]^*\,\det\Big[\Sigma_{{W}}(I,J)\Big]\,\det\big[{\rho}({J,L})\big]\,D_L^{(W)}({z})\\
&\,=D_K^{(W)}(\bar{z})\,\det\big[{\rho}(I,K)^*\Sigma_{{W}}(I,J){\rho}({J,L})\big]\,D_L^{(W)}({z})\\
&\,=D_K^{(W)}(\bar{z})\,\det\big[\Sigma_{{W}}(K,L)\big]\,D_L^{(W)}({z})\\
&\,=D_K^{(W)}(\bar{z})\,\big(\widetilde{\Sigma}_{M,N}\big)_{KL}\,D_L^{(W)}({z})\,,
\esp\eeq
where the third step follows from the monodromy invariance of $\Sigma_W$. Hence, eq.~\eqref{eq:Phi_MN_Sigma} is monodromy-invariant.


\subsection{The Basso-Dixon formula}

Equation~\eqref{eq:Phi_MN_Sigma} allows us to write all fishnet integrals as bilinears in the determinants of the periods computed from the ladder integrals. From the BD formula in eq.~\eqref{eq:fishnet_kazakov}, we know that the fishnet integrals can also be expressed as determinants in the ladder integrals, i.e., determinants of bilinears of periods. We now show how the BD formula in $D=2$ dimensions arises from the CY geometry.

In the following $\varepsilon_{a_1\cdots a_M}$ denotes the Levi-Civita tensor of rank $M$, i.e., the totally antisymmetric tensor in $M$ indices. We compute
\begin{align}
\det&\left(\theta_{\bar{z}}^{i-1}\theta_{{z}}^{j-1}\phi_{W}(z)\right)_{1\le i,j\le M} \\
\nonumber&\,=\frac{1}{M!}\varepsilon_{a_1\cdots a_M}\varepsilon_{b_1\cdots b_M}\big[\theta_{\bar{z}}^{a_1-1}\theta_z^{b_1-1}\phi_{W}(z)\big]\cdots \big[\theta_{\bar{z}}^{a_M-1}\theta_z^{b_M-1}\phi_{W}(z)\big]\\
\nonumber&\,=\frac{(-i)^{MW^2}}{M!}\varepsilon_{a_1\cdots a_M}\varepsilon_{b_1\cdots b_M}\big[\theta_{\bar{z}}^{a_1-1}{\Pi}_{W,i_1}(\bar{z})\,\big(\Sigma_{W}\big)_{i_1j_1}\,\theta_z^{b_1-1}{\Pi}_{W,j_1}(z)\big]\times\cdots \\
\nonumber&\,\qquad\times\big[\theta_{\bar{z}}^{a_M-1}{\Pi}_{W,i_M}(\bar{z})\big(\Sigma_{W}\big)_{i_Mj_M}\theta_z^{b_M-1}{\Pi}_{W,j_M}(z)\big]
\end{align}
\begin{align}
\phantom{\det}
\nonumber&\,=\frac{(-i)^{MW^2}}{M!}\big[\varepsilon_{a_1\cdots a_M}\theta_{\bar{z}}^{a_1-1}{\Pi}_{W,_1}(\bar{z})\cdots\theta_{\bar{z}}^{a_M-1}{\Pi}_{W,i_M}(\bar{z})\big]\,\big(\Sigma_{W}\big)_{i_1j_1}\cdots \big(\Sigma_{W}\big)_{i_Mj_M}\times\\
\nonumber&\,\qquad \times\big[\varepsilon_{b_1\cdots b_M}\theta_z^{b_1-1}{\Pi}_{W,j_1}(z)\cdots \theta_z^{b_M-1}{\Pi}_{W,j_M}(z)\big]\,.
\end{align}
We have, with $J = (j_1,\ldots, j_M)$,
\beq\bsp
\varepsilon_{b_1\cdots b_M}\theta_z^{b_1-1}{\Pi}_{W,j_1}(z)\cdots \theta_z^{b_M-1}{\Pi}_{W,j_M}(z) &\,= \varepsilon_{j_1\cdots j_M}\det\big(\theta_z^{b}{\Pi}_{W,j}(z)\big)_{1\le b\le M, j\in J} \\
&\,= \varepsilon_{j_1\cdots j_M}\, {D}^{(W)}_J(z)\,,
\esp\eeq
and similarly for the anti-holomorphic contribution. Hence,\footnote{We stress that a priori the last equality follows only for the lower loop cases in which we checked the determinant form of the Basso-Dixon periods, cf.\ the discussion around \eqref{detDI}. Since we know, however, from ~\cite{Derkachov:2018rot} that the 2d Basso-Dixon formula holds in general, it is clear that this equality also holds in general.}
\begin{align}
\det&\left(\theta_{\bar{z}}^{i-1}\theta_{{z}}^{j-1}\phi_{W}(z)\right)_{1\le i,j\le M} \\
\nonumber&\,=(-i)^{MW^2}\, {D}^{(W)}_I(\bar{z})\,\left[\frac{1}{M!}\varepsilon_{i_1\cdots i_M}\varepsilon_{j_1\cdots j_M}\,\big(\Sigma_{W}\big)_{i_1j_1}\cdots \big(\Sigma_{W}\big)_{i_Mj_M}\right]\,{D}^{(W)}_J(z)\\
\nonumber&\,= (-i)^{MW^2}{D}^{(W)}_{I}(\bar{z})\,\det\Big[\big(\Sigma_{W}\big)_{ij}\Big]_{\substack{i\in I, j\in J}}\,{D}^{(W)}_J(z)\\
\nonumber&\,= (-i)^{MW^2}{D}^{(W)}_{I}(\bar{z})\,\big(\widetilde\Sigma_{M,N}\big)_{IJ}\,{D}^{(W)}_J(z)\\
\nonumber&\,=(-i)^{MW^2}\,i^{(MN)^2}\phi_{M,N}(z)\,.
\end{align}
To summarize, we see that we have the relation:
\beq\label{eq:our_BD}
\phi_{M,N}(z) =(-i)^{(MN)^2-M(M+N-1)^2}\, \det\left(\theta_{\bar{z}}^{i-1}\theta_{{z}}^{j-1}\phi_{M+N-1}(z)\right)_{1\le i,j\le M}\,.
\eeq
The previous relation allows one to write, up to an overall constant factor, the fishnet integral $\phi_{M,N}(z)$ as an $M\times M$ determinant involving the derivatives of the ladder integral with $W=M+N-1$ loops. It is in fact identical to the BD-formula for 2D fishnet integrals of ref.~\cite{Derkachov:2018rot}, see eq.~\eqref{eq:fishnet_kazakov}. The only apparent differences come from our normalisation of the ladder integrals and the choice of the conformal cross ratio. 

Equation~\eqref{eq:our_BD} shows that the BD formula for $\phi_{M,N}(z)$ in two dimensions is in fact  equivalent to the statement that $L_{M,N}=\wedge^ML_W$. The latter statement, however, has only been checked explicitly for low loop orders, and it still remains conjectural in the general case. However, since the BD formula was shown to hold independently in ref.~\cite{Derkachov:2018rot}, this shows that $L_{M,N}=\wedge^ML_W$ holds for all values of $(M,N)$.

%% file: volumes.tex

  \section{Volumes and iterated integrals}
  \label{sec:volumes}
  In this section we argue that our multi-valued definition of the quantum volume in eq.~\eqref{eq:Vol_q_def} has an interesting feature, which makes it particularly appealing in the context of Feynman integrals, namely we expect that $\textrm{Vol}_{\textrm{q}}(\mathcal{W}_{M,N})$ can be written in terms of iterated integrals with at most logarithmic singularities at the MUM-point.\footnote{For the calibrated volume of $\mathcal{M}_{M,N}$, the statement remains true after factoring out the period that is holomorphic at the MUM point.} Such iterated integrals appear frequently in the context of Feynman integrals, where they are often referred to as \emph{pure functions}~\cite{ArkaniHamed:2010gh,Broedel:2018qkq}. While we only show this explicitly for the case of CY operators and their exterior powers (which cover all four-point fishnet integrals considered in this paper), we strongly expect this feature to persist in other cases. 

Let us first discuss the case of the ladder integrals $\phi_N(z) = \phi_{1,N}(z)$. We have seen that the Picard-Fuchs operator $L_N$ is a CY operator of degree $N+1$. The mirror map becomes (cf.~eq.~\eqref{eq:mirror_map})
\beq\label{eq:1-parameter-q}
t(z) = \frac{1}{2\pi i}\,\frac{\Pi_{N,1}(z)}{\Pi_{N,0}(z)} = \frac{\log z}{2\pi i} + \mathcal{O}(z)\,.
\eeq
In ref.~\cite{Duhr:2022dxb} it was shown that one can write the (normalised) solutions of a CY operator in terms of iterated integrals ($k<N$):
\beq\label{eq:Pi_to_I}
\Pi_{N,k}(z(q)) = \Pi_{N,0}(z(q)) \,I(Y_{N,0},Y_{N,1},\dots Y_{N,k-1};q) = \frac{1}{k!}\log^k q + \mathcal{O}(q)\,,\qquad q:=e^{2\pi i t(z)}\,,
\eeq
where we defined the iterated integral~\cite{ChenSymbol}:\footnote{Note that this integral may exhibit a divergence at $q=0$, and we need to replace it by a suitably regularised version, e.g., by interpreting the lower integration boundary as a tangential base-point. We refer to ref.~\cite{Brown:mmv} for a detailed discussion.}
\beq
I(f_1,\ldots,f_k;q) := \int_0^q \frac{\rd q'}{q'}\,f_1(q')\,I(f_2,\ldots,f_k;q)\,,\qquad I(;q)=1\,.
\eeq
The arguments of the iterated integrals are the \emph{$Y$-invariants} attached to the CY operator $L$~\cite{BognerCY,BognerThesis}, which can be defined in terms of the structure series $\alpha_m(z)$ defined in section~\ref{sec:CY_ops}:
\beq
\label{Yinvariants}
Y_{N,0}(q) = 1 \textrm{~~~and~~~}Y_{N,k}(q) = \frac{\alpha_{1}(z(q))}{\alpha_{k+1}(z(q))}\,,  \quad \text{for } 1\le k\le N-2\,.
\eeq
The $Y$-invariants satisfy the identity:
\beq\label{eq:Y_ref}
Y_{N,k}(q) = Y_{N,N-1-k}(q)\,.
\eeq
Inserting eq.~\eqref{eq:Pi_to_I} into eq.~\eqref{eq:Vol_q_def}, we obtain:
\beq
\textrm{Vol}_{\textrm{q}}(\mathcal{W}_{{1,N}}) = {\underline{I}_{1,N}(q)^{\dagger}\Sigma_N\underline{I}_{1,N}(q)}\, ,
\eeq
where $\underline{I}_{1,N}(q)$ is the vector of iterated integrals:
\beq\bsp\label{eq:I1N_def}
\underline{I}_{1,N}(q)&\,= \big(1,I(Y_{N,0};q), I(Y_{N,0},Y_{N,1};q),\ldots, I(Y_{N,0},\ldots,Y_{N,N-1};q)\big)^T \\
&\,= \big(\delta^{N-k}I(Y_{N,0},\ldots,Y_{N,N-1};q)\big)_{0\le k\le N}\,,
\esp\eeq
where we introduced the derivation\footnote{It is well-known that iterated integrals form a shuffle algebra under multiplication. The operation $\delta$ is a genuine derivation for the shuffle product, i.e., it is linear and satisfies the Leibniz rule.} $\delta$ that acts on iterated integrals by clipping off letters from the right:
\beq\label{eq:delta_def}
\delta I(f_1,\ldots,f_k;q) := I(f_1,\ldots,f_{k-1};q)\textrm{~~~and~~~} \delta(1) = 0\,.
\eeq
We see that the quantum volume for a ladder integral can be written in terms of iterated integrals with only logarithmic singularities at the MUM-point. Note that this is not restricted to ladder integrals, but exactly the same argument applies to a quantum volume computed from any CY operator.

Let us now turn to fishnet integrals $\phi_{M,N}(z)$ with $M>1$. We have seen that the solution space of the Picard-Fuchs operator is generated by the determinants in eq.~\eqref{eq:det_periods}. Since the determinants also involve the holomorphic period ${\Pi}_{{W},0}(z)$ (with ${W}=M+N-1$), it is not entirely obvious that the quantum volume can be written in terms of iterated integrals with only logarithmic singularities at the MUM-point. In appendix~\ref{app:proof_IIQ} we show that the following identity holds:
\beq\label{eq:det_to_Delta}
{D}_I^{({W})}(z) = {D}_{0\cdots (M-1)}^{({W})}(z)\,\Delta_{M,N,I}({q})\,,
\eeq
where we defined:
\beq\label{eq:cI_Def}
 \Delta_{M,N,I}({q}) = \det\left(\delta^{{W}-i}\vartheta^jI(Y_{W,0},\ldots,Y_{W,W-1};q)\right)_{\substack{i\in I\\ 0\le j<M}}
\,.
\eeq
Here $\vartheta$ denotes the derivation that clips off letters from the left (cf.~eq.~\eqref{eq:delta_def}):
\beq\label{eq:vartheta_def}
\vartheta I(f_1,\ldots,f_k;q) := I(f_2,\ldots,f_{k};q)\textrm{~~~and~~~} \vartheta(1) = 0\,.
\eeq
Note that $\vartheta$ is related to the ordinary derivative $\theta_q$ by
\beq
\theta_qI(f_1,\ldots,f_k;q) = f_1(q)\,\vartheta I(f_1,\ldots,f_k;q)\,.
\eeq
%
The argument $q$ of the iterated integrals in eq.~\eqref{eq:cI_Def} is the canonical $q$-coordinate of the CY operator $L_{{W}}$, cf.~eqs.~\eqref{eq:1-parameter-q} and~\eqref{eq:Pi_to_I}. We would expect, however, that the natural variable in the context of the operators $L_{M,N}$ is 
\beq\label{eq:t_tilde_def1}
\tilde{t}(z) = \frac{1}{2\pi i}\,\frac{{\Pi}_{M,N,1}(z)}{{\Pi}_{M,N,0}(z)}\,,\qquad \tilde{q} = e^{2\pi i \tilde{t}(z)}\,.
\eeq
In appendix~\ref{app:proof} we show that ${D}_{0\cdots (M-1)}^{({W})}(z)$ is always holomorphic at the MUM-point $z=0$, and ${D}_{0\cdots (M-2)M}^{({W})}(z)$ always diverges like $\log z$ as $z\to0$. Hence, we have ${\Pi}_{M,N,0}(z) = {D}_{0\cdots (M-1)}^{({W})}(z)$ and ${\Pi}_{M,N,1}(z) = {D}_{0\cdots (M-2)M}^{({W})}(z)$. Using eqs.~\eqref{eq:det_to_Delta} and~\eqref{eq:cI_Def}, we find the following very simple relation between $t(z)$ and $\tilde{t}(z)$:
\beq\label{eq:t_tilde_def}
\tilde{t}(z) = \frac{1}{2\pi i}\,I(Y_{{W},M-1};q)\,,
\eeq
and so
\beq
\tilde{q} = \exp\left[I(Y_{{W},M-1};q)\right] = q + \mathcal{O}(q^2)\,.
\eeq
It is then easy to show that we have
\beq
I({Y}_{{W},i_1},\ldots,{Y}_{{W},i_k};{q}) = I(\widetilde{Y}_{M,{W},i_1},\ldots,\widetilde{Y}_{M,{W},i_k};\tilde{q}) \,,
\eeq
where we defined
\beq
\widetilde{Y}_{M,{W},i}(\tilde{q}) \coloneqq \frac{{Y}_{{W},i}(q)}{{Y}_{{W},M-1}(q)}\,.
\eeq
Putting everything together, we obtain
\beq
\textrm{Vol}_{\textrm{q}}(\mathcal{W}_{M,N}) = {\underline{I}_{M,N}(\tilde{q})^{\dagger}\widetilde{\Sigma}_{M,N}\underline{I}_{M,N}(\tilde{q})}\,,
\eeq
where $\Sigma_{M,N}$ was defined in eq.~\eqref{eq:Sigma_M_N_def} and $\underline{I}_{M,N}(\tilde{q})$ is the vector of iterated integrals:
\beq
\underline{I}_{M,N}(\tilde{q})= \big(\Delta_{M,N,I}(\tilde{q})\big)_{I\in \overline{T}_{W,M}}^T\,.
\eeq
To conclude, we see that in all cases $\textrm{Vol}_{\textrm{q}}(\mathcal{W}_{M,N})$ can be expressed as a linear combination of iterated integrals with only logarithmic singularities at the cusp. The arguments of the iterated integrals are the $Y$-invariants of the ${W}$-loop ladder $\phi_{W}(z)$. 

Finally, we can repeat exactly the same steps as in the derivation of the BD-formula in eq.~\eqref{eq:our_BD} (with the derivative $\theta_z$ replaced by the derivation $\vartheta$) to obtain a variant of the BD-formula directly for the quantum volume of $\mathcal{W}_{{M,N}}$:
\beq
\textrm{Vol}_{\textrm{q}}(\mathcal{W}_{M,N}) = \det\left[\bar{\vartheta}^{i-1}\vartheta^{j-1}\textrm{Vol}_{\textrm{q}}(\mathcal{W}_{1,M+N-1}) \right]_{0\le i,j<M}\,,
\eeq
where $\bar{\vartheta}$ denotes the analogue of $\vartheta$, but acting on anti-holomorphic iterated integrals.

%% file: app_proof_det.tex

\section{Exterior powers of Calabi-Yau operators} 
\label{app:proof}

In this appendix we show that, if $L_W$ is a CY operator of degree $W+1$, then its exterior powers $\wedge^ML_W$ are differential operators with a solution space structure expected a Picard-Fuchs operator for a CY $\ell$-fold, with $\ell=MN$ and $N=W-M+1$. Throughout this section, we will use the notations that we introduced for ladder and fishnet integrals, but all results apply to arbitrary CY operators. We note that, for arbitrary CY operators, the results of this section do not prove that there is a family of CY $\ell$-folds with Picard-Fuchs operator is $\wedge^ML_W$. In the case of fishnet integrals, however, we know that it is the Picard-Fuchs operator of the family $\mathcal{M}_{M,N}$.

We have seen in section~\ref{sec:DEQ_ops} that a spanning set for $\Sol(\wedge^ML_W)$ is given by the determinants ${D}_I^{(W)}(z)$ in eq.~\eqref{eq:det_form}, where $I=\{i_1,\ldots,i_M\}\subseteq \{1,\ldots,W\}$ is an ordered set. It is easy to check that $\wedge^ML_W$ has a MUM-point at $z=0$. In the following we show that the number $h_p$ of ${D}_I^{(W)}(z)$ that behave like $\log^pz$ as $z\to 0$ has the following properties:
\beq\bsp\label{eq:h_structure_claim}
h_0&\,=h_1= h_{\ell} = h_{\ell-1}= 1\,,\\
h_{p}&\,=0\,, \textrm{~~if } p>\ell\,,\\
h_{\ell-p} &\,=h_{p}\,, \textrm{~~for all } 0\le p\le \ell\,.
\esp\eeq
This is precisely the structure of the Hodge numbers $h^{p,\ell-p}$ that one expects from a one-parameter family of CY $\ell$-folds. Moreover, we will show that the determinants behave in the limit $z\to 0$ as follows:
\begin{align}
\label{det.3}
{D}_I^{(W)}(z) \sim \log^{n_{W,I}}z\text{ with } n_{W,I} = \sum_{i\in I} i -  \frac{(M-1)M}{2} \, . 
\end{align}
This implies that we can identify uniquely the determinants that behave like $\log^pz$,  $p\in\{0,1,\ell-1,\ell\}$: 
\beq\bsp\label{eq:4Ds}
{D}_{01\dots(M-1)}^{(W)}(z) &\,\sim \log^0z \textrm{~~for } p=0\, ,\\
{D}_{01\dots(M-2)M}^{(W)}(z) &\,\sim \log^1z \textrm{~~for } p=1\, ,\\
{D}_{(W-M)(W-M+2)\ldots W}^{(W)}(z) &\,\sim \log^{\ell-1}z \textrm{~~for } p=\ell-1\, ,\\
{D}_{(W-M+1)\ldots W}^{(W)}(z) &\,\sim \log^{\ell}z \textrm{~~for } p=\ell\,.
\esp\eeq
In particular, we can uniquely identify the solutions that are holomorphic or diverge like a single power of $\log z$ at the MUM-point, and so we can write down the mirror map for $\wedge^ML_W$ (cf.~eq.~\eqref{eq:t_tilde_def1}):
\beq
\tilde{t}(z) = \frac{1}{2\pi i}\,\frac{{D}_{01\dots(M-2)M}^{(W)}(z)}{{D}_{01\dots(M-1)}^{(W)}(z)}\,.
\eeq
In the remainder  of this appendix we present the proofs of these claims.

Since $L_W$ is assumed to be a CY operator, we know that its solution space admits a basis of the form
\begin{align}\label{det.1}
{\Pi}_{W,k} (z) =\sum_{j=0}^k \frac{1}{(k-j)!}\mySigma_{j}(z)\, \log^{k-j} z \, ,
\end{align}
where the $\mySigma_{k}(z)$ are holomorphic in a neighborhood of the MUM-point $z=0$ and normalized such that $\mySigma_{k}(z) = \delta_{k0}\,z+\mathcal{O}(z^2)$.

%
%
%
%
%
We compute $D_I^{(W)}(z) $ with the Leibniz determinant formula: 
\beq\bsp
&D_I^{(W)}(z) = \varepsilon_{j_1,\dots, j_M} \left(\partial_z^{j_1-1}\Pi_{W,i_1} (z)\right) \left(\partial_z^{j_2-2}\Pi_{W,i_2} (z)\right) \dots  \left(\partial_z^{j_M-1}\Pi_{W,i_M} (z)\right)\notag\\
 &=\varepsilon_{j_1,\dots, j_M} \prod_{r=1}^{M}\sum_{k_r=0}^{i_r} \frac{1}{(i_r-k_r)!}\left(\partial_z^{j_r-1} \log^{i_r-k_r}(z) \mySigma_{k_r}(z)\right)\label{det.20}\, .
\esp\eeq
Then, we use the Leibniz product rule to express eq.~\eqref{det.20} as
 \begin{align}\label{det.10} 
&D_I^{(W)}(z) =\varepsilon_{j_1,\dots, j_M} \prod_{r=1}^{M}\sum_{k_r=0}^{i_r}\sum_{t_r=0}^{j_r-1} \begin{pmatrix} j_r-1\\t_r\end{pmatrix}\frac{1}{(i_r-k_r)!}\left(\partial_z^{t_r} \log^{i_r-k_r}z \right) \partial_{z}^{j_{r}-1-t_r} \mySigma_{k_r}(z)\, .
\end{align}
 Naively, one might assume that the term with the highest power of $\log z$ in eq.~\eqref{det.10} corresponds to the term with $k_r=0$ and $t_r=0$ for all $r$. Explicitly, this term is  
\begin{align*}
\varepsilon_{j_1,\dots, j_M} \prod_{r=1}^W  \frac{1}{i_r!}\log^{i_r}z\, \partial_z^{j_r-1} \mySigma_{0}(z)\, . 
\end{align*}
The product in this term is totally symmetric under exchanges of the $j_r$, while the Levi-Civita tensor $\varepsilon_{j_1,\dots, j_M}$  is by definition totally antisymmetric under these exchanges. So,  this term vanishes, and there are no contributions from terms of the form $\log^{\sum_{r=1}^M i_r}z$ to $D_I^{(W)}(z) $. Similar cancellations also happen for terms of lower powers of $\log z$, and it turns out that the highest appearing power depends on $M$ and $W$. 

 To see this, we first consider each factor of the product in eq.~\eqref{det.10} separately. Let $C_{j_r}^{i_r-k}$ be the sum of all terms of order $\log^{i_r-k}z$  from the $r$-th factor in the product, for example: 
\begin{align}
C_{j_r}^{i_r}(z) &=   \frac{1}{i_r!}\log^{i_r}z\, \partial_z^{j_r-1} \mySigma_{0}\,, \\
C_{j_r}^{i_r-1}(z) &=\sum_{t_r=1}^{j_r-1}\Bigg[ \begin{pmatrix}j_r-1\\t_r\end{pmatrix} \frac{1}{i_r!}
\left(\partial_z^{t_r} \log^{i_r}z \right) \Big|_{\log^{i_r-1}z}\partial_{z}^{j_{r}-1-t_r} \mySigma_{0}(z)+   \frac{\log^{i_r-1}z }{(i_r-1)!} \partial_{z}^{j_{r}-1} \mySigma_{1}(z)\Bigg]\notag\\
&=\left[\sum_{t_r=1}^{j_r-1} H_{i_r-1}^{t_r}(z)\begin{pmatrix}j_r-1\\t_r\end{pmatrix}\partial_{z}^{j_{r}-1-t_r} \mySigma_{0}(z)+  \partial_{z}^{j_{r}-1} \mySigma_{1}(z)\right]\frac{1}{(i-1)!}\log^{i_r-1}z \,.
\end{align}
The factor $H_{i_r-1}^{t_r}(z)$ represents the coefficient of $\log^{i_r-1} z$ in $\partial_{z}^{t_r}\log^{i_r}z$, divided by $i_r$. For example, we have:
\begin{align*}
\partial_z \log^{i_r} z &= i_r \frac{\log^{i_r-1}z}{z} \\
&\rightarrow H_{i_r-1}^{1} (z) =\frac{1}{z}\\
\partial^2_z \log^{i_r}z &= i_r(i_r-1) \frac{\log^{i_r-2}z}{z^2}- i_r \frac{\log^{i_r-1}z}{z^2} \\
& \rightarrow H_{i_r-1}^{2} (z) =-\frac{1}{z^2}\\
\partial^3_z \log^{i_r} z &= i_r(i_r-1)(i_r-2) \frac{\log^{i_r-3}z}{z^3}-3i_r(i_r-1) \frac{\log^{i_r-2}z}{z^3} + 2 i_r \frac{\log^{i_r-1}z}{z^3} \\
& \rightarrow H_{i_r-1}^{3} (z) =\frac{2}{z^2}\, . 
\end{align*}
Similarly, the $C_{j_r}^{i_r-k}(z)$ with $k>1$ are given by:
 \begin{align}
 C_{j_r}^{i_r-k}(z)
 &=\Bigg[\sum_{t_r=k}^{j_r-1} H_{i_r-k,k}^{t_r}(z)\begin{pmatrix}j_r-1\\t_r\end{pmatrix}\partial_{z}^{j_{r}-1-t_r} \mySigma_{0}(z)+\dots\label{det.13}\\ 
 &+\sum_{t_r=1}^{j_r-1} H_{i_r-k,1}^{t_r}(z)\begin{pmatrix}j_r-1\\t_r\end{pmatrix}\partial_{z}^{j_{r}-1-t_r} \mySigma_{k-1}(z)
 +  \partial_{z}^{j_{r}-1} \mySigma_{k}(z)\Bigg]  \times\frac{\log^{i_r-k}z }{(i_r-k)!}
 \, . \notag
\end{align}
We now observe that each of the factors $H_{i_r-k, s}^{t_r}(z)$ is independent of $i_r$, which implies that each $C_{j_r}^{i_r-k}(z)$ can be written as a product of the form
\begin{align}
C_{j_r}^{i_r-k}(z)&= F_{j_r}^k(z) \times  \frac{\log^{i_r-k}z}{(i_r-k)!}
  \, , 
\end{align}
where $F^k_{j_r}(z)$ is defined by eq.~\eqref{det.13} and only depends on $j_r$ and $k$. In this form, it is obvious that a product of two or more $C_{j_r}^{i_r-k}(z)$ with the same $k$ is invariant under exchanges of the $j_r$:
\begin{align}
C_{j_r}^{i_r-k}(z) C_{j_s}^{i_s-k}(z) = F_{j_r}^k(z) F_{j_s}^k(z)    
  \frac{\log^{i_r-k}z}{(i_r-k)!}\frac{\log^{i_s-k}z}{(i_s-k)!}
= C_{j_s}^{i_r-k}(z) C_{j_r}^{i_s-k}(z)\, .
\end{align}
All terms in eq.~\eqref{det.10} containing such symmetric products vanish due to the presence of the Levi-Civita tensor.

 The total power of $\log z$ in a product  $\prod_{r=1}^{m}C_{r,j_r}^{i_r-k_r}(z)$ is $ \sum_{i\in I}( i-k_i)$. We just saw that if any two of the $k_i$ in such a product are equal, the corresponding term in $D_I^{(W)} $ vanishes. So, the highest non-vanishing power of $\log z$ corresponds to the smallest set of distinct $k_i$ -- which is any permutation of the set $ \{0,1,\dots, p-1\}$. 
Thus the terms with the highest order in $\log z$  have the form
\begin{align}
\label{det.8}
C_{j_1}^{i_1-0}(z)C_{j_2}^{i_r-1}(z) \dots C_{j_M}^{i_M-(M-1)}(z) \sim \big(\log z\big)^{\sum_{i\in I}i -\sum_{j=0}^{M-1} j} = \big(\log z\big)^{\sum_{i\in I}i -\frac{M(M-1)}{2}}\, . 
\end{align}
We can assume this product is actually non-vanishing for general $\mySigma_{k}$, because the terms in it do not  cancel each other: Any $C_{j_r}^{i_r-k}$ contains a term $\frac{1}{(i_r-k)!} \log^{i_r-k}z\,  \partial_{z}^{j_{r}-1} \mySigma_{k}(z)$, which means that the product in eq.~\eqref{det.8}
always contains a term of the form:
\begin{align}
\label{det.15}
\frac{1}{i_r!}\log^{i_r}z\,  \partial_{z}^{j_{r}-1} \mySigma_{0}(z)\times\dots \times \frac{1}{(i_r-(M-1))!}\log^{i_r-(M-1)}z \, \partial_{z}^{j_{r}-1} \mySigma_{M-1}(z)\, .
\end{align}
 Being the product of the terms with the highest $k_r$ in each of the $C_{j_r}^{i_r-k}(z)$, it is the only one containing the product $\partial_{z}^{j_r-1}\mySigma_{0}(z)\dots \partial_z^{j_r-1}\mySigma_{M-1}(z)$ and thus cannot cancel with any of the other terms for general $\mySigma_{k}$.
This proves the claim in eq.~\eqref{det.3}. 

Equation~\eqref{eq:h_structure_claim} and~\eqref{eq:4Ds} now follow from eq.~\eqref{det.3}. We  assume $M$ and $\ell$ to be fixed for the rest of this appendix. Let us start by discussing the cases of solutions that behave like $\log^pz$, with $p=0,1$:
\begin{itemize}
\item For $D_I^{(W)} \sim \log^0z$, the subset $I$ must fulfil: $0=n^{\text{min}}_{W,I} = \sum_{i\in I} i-\sum_{j=0}^{M-1}j$. This means, $I$ can only be $\{0,1,\dots , M-1\}$, since we only allow ordered subsets $I$ with distinct entries $i_j$. Hence the unique holomorphic solution is $D_{01\ldots(M-1)}^{(W)}$, and so $h_0=1$.
\item For $D_I^{(W)} \sim \log^1z$, the subset $I$ must fulfil: $1=n_{W,I} = \sum_{i\in I}i -\sum_{j=0}^{M-1}j$. Again, since we have distinct and ordered entries in $I$, the only possible $I$ that fulfils this condition is $\{0,1,\dots, M-2,M\}$. If we had changed any other entry of $I$ from the first case, we would also need to change another one to have distinct entries. But then $n_{W,I}$ would be larger than $1$. Hence the unique solution that behaves like $\log z$ is $D_{01\ldots(M-2)M}^{(W)}$, and so $h_1=1$.
\end{itemize}
Very similarly, we see: 
\begin{itemize}
\item For the $D_I^{(W)} $ with the maximal value of $n_{W,I} = \sum_{i\in I} i-\frac{M(M-1)}{2}$, we need to have the largest possible entries in $I$. This means, $I$ must be $\{W-M+1,\dots, W-1,W\}$. In that case we have
\begin{align}
n^{\text{max}}_{W,I} &= \sum_{i=W-M}^{W} i - \sum_{j=0}^{M-1} j 
= \sum_{k=0}^{M-1} (W-M+1)
= M(W-M+1)\, . 
\end{align}
Thus, there is no solution that behaves like $\log^pz$, $p>M(W-M+1)$, and so $h_p=0$ in that case. The  unique solution that behaves like $\big(\log z\big)^{M(W-M+1)}$ is  $D_{(W-M+1)\dots(W-1)W}^{(W)}$, and so $h_{M(W-M+1)}=1$. 
\item By a very similar argument, we observe that the only solution that behaves like $\big(\log z\big)^{M(W-M+1)-1}$ is For  $D_{(W-M)(W-M+2)\dots W}^{(W)}$, and so $h_{M(W-M+1)-1}=1$. 
\end{itemize}

Finally, we need to show that we have the symmetry property $h_{M(W-M+1)-p}=h_p$. This is equivalent to showing that, for fixed $W$ and $M$, we get the same number of distinct $I$ with  $D_I^{(W)}$ that behave like $\log^{n^{\text{min}}_{W,I}+x}z= \log^{x}z$ and $\log^{n^{\text{max}}_{W,I} -x}$. To see that this is indeed true, consider  the function 
\begin{align}
J_{W}(y,x)= \prod_{j=0}^W (1+yx^j)= \sum_{M=0}^{W} \sum^{\frac{M(2W-M+1)}{2}}_{S=\frac{M(M-1)}{2}} J^{M,S}_{W}y^M x^S \, . 
\end{align}
The term $J_N^{M,S} y^M x^S$ is a sum of all terms that are of the form $yx^{j_1}\dots yx^{j_p}= y^M x^S$, products of $M$ factors $yx^{j_s}$ such that the $j_s$ sum to $S$. Thus, the coefficient $J^{M,S}_N$ is the number of sets $I=\{i_1,\dots, i_M\}$ of $\{0,\dots, W\}$ such that $\sum_{k=1}^{M}i_k=S$. For $S_{\text{min}}= \frac{M(M-1)}{2}$ and $S_{\text{max}} =\frac{M(2W-M+1)}{2}$: 
\begin{align}
J_W^{M, S_{\text{min}}+x}=J_W^{M, S_{\text{max}}-x} \, . 
\end{align}
The number of distinct $I$ that sum to $S_{\text{min}}+x$ and $S_{\text{max}}-x$ is equal, which is equivalent to the number of distinct $I$ with $D_I^{(W)} \sim \log^{n^{\text{min}}_{W,I}+x}z= \log^{x}z$ and with $D_I^{(W)} \sim \log^{n^{\text{max}}_{W,I} -x}z$ being equal.

%% file: app_proof_rels.tex

\section{Linear relations among determinants of periods} 
\label{app:proof_rels}

In this section we proof that for $W$ odd and $M\le\frac{W-1}{2}$, the $M\times M$ determinants $D^{(W)}_I(z)$ satisfy the linear relations in eq.~\eqref{eq:det_relations}. We follow our notations and conventions for fishnet graphs, but we emphasise that the conclusions hold for an arbitrary CY operator $L_W$ of even degree $M+N=W+1$.

The periods of a family of CY varieties and their derivatives satisfy the quadratic relations in eq.~\eqref{Griffiths}. They imply the following relations:
\beq\label{eq:Griffith}
\theta_z^a\uPi_W(z)^T\Sigma_W\theta_z^b\uPi_W(z) = 0\,,\textrm{~~if~} a+b<W\,.
\eeq
If $W$ is odd, then $\Sigma_W$ is antisymmetric, and it is easy to show that we have:
\beq
\theta_z^a\uPi_W(z)^T\Sigma_W\theta_z^b\uPi_W(z) = \sum_{k=0}^{\frac{W-1}{2}}(-1)^k\,\Delta_k^{ab}(z)\,,
\eeq
where we defined
\beq
\Delta_k^{ab}(z) = \det\left(\begin{matrix}\theta_z^a\Pi_{W,k}(z) & \theta_z^a\Pi_{W,W-k}(z)\\
\theta_z^b\Pi_{W,k}(z) & \theta_z^b\Pi_{W,W-k}(z)\end{matrix}\right)\,.
\eeq

Let $J=(j_1,\ldots,j_{M-2})$. We define 
\beq
J_k = (k,W-k,j_1,\ldots,j_{M-1})\,,\qquad 0\le k\le \frac{W-1}{2}\,.
\eeq
The left-hand side of eq.~\eqref{eq:det_relations} can then be written in the form
\beq\bsp
\sum_{k=0}^{\frac{W-1}{2}}&(-1)^k\,D_{J_k}^{(W)}(z) =\\
&\,=\sum_{k=0}^{\frac{W-1}{2}}\sum_{0\le a_i <M}\sum_{b_i\in J_k}\frac{(-1)^k}{M!}\varepsilon_{a_1\cdots a_M}\varepsilon_{b_1\cdots b_M}\theta_z^{a_1}\Pi_{W,b_1}(z)\ldots \theta_z^{a_M}\Pi_{W,b_M}(z)\,.
\esp\eeq
We now focus on those terms where the elements of $J$ appear in a fixed position inside $(b_1,\ldots,b_M)$. We only discuss the case $(b_3,\ldots,b_M)=(j_1,\ldots,j_{M-2})$. All other cases are similar. 

The terms with $(b_3,\ldots,b_M)=(j_1,\ldots,j_{M-2})$ can be cast in the form (up to an irrelevant overall factor):
\beq\bsp
\sum_{0\le a_i <M}&\sum_{k=0}^{\frac{M-1}{2}}\frac{(-1)^k}{M!}\varepsilon_{a_1\cdots a_M}\theta_z^{a_3}\Pi_{W,j_1}(z)\ldots \theta_z^{a_M}\Pi_{W,j_{M-2}}(z)\,\times\\
&\,\qquad\times\Big[\theta_z^{a_1}\Pi_{W,k}(z)\theta_z^{a_2}\Pi_{W,W-k}(z)-\theta_z^{a_1}\Pi_{W,W-k}(z)\theta_z^{a_2}\Pi_{W,k}(z)\Big]\\
&\,=\sum_{0\le a_i <M}\frac{1}{M!}\varepsilon_{a_1\cdots a_M}\theta_z^{a_3}\Pi_{W,j_1}(z)\ldots \theta_z^{a_M}\Pi_{W,j_{M-2}}(z)\,\sum_{k=0}^{\frac{M-1}{2}}{(-1)^k}\Delta^{a_1a_2}_k(z)\\
&\,=\sum_{0\le a_i <M}\frac{1}{M!}\varepsilon_{a_1\cdots a_M}\theta_z^{a_3}\Pi_{W,j_1}(z)\ldots \theta_z^{a_M}\Pi_{W,j_{M-2}}(z)\,\left[\theta_z^{a_1}\uPi_W(z)^T\Sigma_W\theta_z^{a_2}\uPi_W(z)\right]\\
&\,=0\,,
\esp\eeq
where the last step follows from eq.~\eqref{eq:Griffith}, because, for $M\le\frac{W-1}{2}$, we have:
\beq
a_1+a_2\le 2(M-1)\le W-3<W\,.
\eeq

%% file: app_proof_IIQ.tex

\section{From determinants to iterated integrals}
\label{app:proof_IIQ}

In this appendix, we present the proof of eq.~\eqref{eq:det_to_Delta}. For $I=(i_1,\ldots,i_M)$ (with $i_p<i_{p+1}$) we can write
\beq\label{eq:appB_eq_0}
D^{(W)}(z) = \det\left(\uPi_{I}(z), \theta_z\uPi_{{I}}(z),\ldots,\theta_z^M\Pi_{{I}}(z)\right)\,,
\eeq 
where we introduced the shorthand
\beq
\uPi_{I}(z):=\left(\uPi_{{W,i_1}}(z),\ldots,\uPi_{{W,i_M}}(z)\right)^T\,.
\eeq
For the first column we have
\beq
\uPi_{{I}}(z) = \Pi_{W,0}(z)\,\underline{I}_{{I}}(q)\,,
\eeq
where we defined 
\beq
\underline{I}_{{I}}(q):=\left(I(Y_{W,0},\ldots,Y_{W,i_1-1};q),\ldots,I(Y_{W,0},\ldots,Y_{W,i_M-1};q)\right)^T\,.
\eeq
The overall factor $\Pi_{W,0}(z)$ can be factored out of the determinant. The second column can be written as
\beq\bsp\label{eq:appB_eq_11}
\theta_z\uPi_{{I}}(z) &\,= \theta_z\left(\Pi_{W,0}(z)\,\underline{I}_{{I}}(q)\right)=\underline{I}_{{I}}(q)\,\theta_z\Pi_{W,0}(z)+\Pi_{W,0}(z)\,\theta_z\underline{I}_{{I}}(q)\,.
\esp\eeq
The first term in eq.~\eqref{eq:appB_eq_11} is proportional to the first column in eq.~\eqref{eq:appB_eq_0}, and so it can be neglected when computing the determinant. The third column in eq.~\eqref{eq:appB_eq_0} can be written as
\beq\bsp\label{eq:appB_eq_1}
\theta_z^2\uPi_{{I}}(z) &\,= \theta_z^2\left(\Pi_{W,0}(z)\,\underline{I}_{{I}}(q)\right)\\
&\,=\underline{I}_{{I}}(q)\,\theta_z^2\Pi_{W,0}(z)+2\theta_z\Pi_{W,0}(z)\,\theta_z\underline{I}_{{I}}(q)+\Pi_{W,0}(z)\,\theta_z^2\underline{I}_{{I}}(q)\,.
\esp\eeq
The first two terms are proportional to the first two columns in eq.~\eqref{eq:appB_eq_0}, and so they can be neglected when computing the determinant. If we continue this way, we arrive at
\beq\label{eq:appB_eq_2}
D_I^{(W)}(z) = \Pi_{W,0}(z)^M\det\left(\underline{I}_{{I}}(q), \theta_z\underline{I}_{{I}}(q),\ldots,\theta_z^M\underline{I}_{{I}}(q)\right)\,.
\eeq 
To continue, we compute
\beq\bsp
\theta_zI(Y_{W,s_1},\ldots,Y_{W,s_l};q) &\,=\frac{1}{\alpha_1(z)}\,\theta_qI(Y_{W,s_1},\ldots,Y_{W,s_l};q)\\
&\,=\frac{1}{\alpha_{s_1+1}(z)}\,I(Y_{s_2},\ldots,Y_{W,s_l};q)\,.
\esp\eeq
From this we obtain for the second column of eq.~\eqref{eq:appB_eq_2}:
\beq
\theta_z\underline{I}_{{I}}(q) = \frac{1}{\alpha_1(z)}\vartheta\underline{I}_{{I}}(q)\,,
\eeq
where the action of the derivation $\vartheta$ on iterated integrals was defined in eq.~\eqref{eq:vartheta_def}, and we extended its action on vectors to be componentwise.
For the third column of eq.~\eqref{eq:appB_eq_2}, we have:
\beq\bsp
\theta_z^2\underline{I}_{{I}}(q) &\,= \theta_z\left(\frac{1}{\alpha_1(z)}\vartheta\underline{I}_{{I}}(q)\right)\\
&\,=-\frac{\theta_z\alpha_1(z)}{\alpha_1(z)^2}\,\vartheta\underline{I}_{{I}}(q) + \frac{1}{\alpha_1(z)\alpha_2(z)}\,\vartheta^2\underline{I}_{{I}}(q)\,.
\esp\eeq
The first term is proportional to the second column, so it can be neglected when computing the determinant. Continuing like this, we arrive at
\begin{align}\label{eq:appB_eq_3}
\nonumber
D_I^{(W)}(z) &\,= \Pi_{W,0}(z)^M\det\left(\underline{I}_{{I}}(q),  {\alpha_1(z)^{-1}}\vartheta\underline{I}_{{I}}(q),\ldots,{\alpha_1(z)^{-1}}\cdots{\alpha_{M-1}(z)^{-1}}\vartheta^{M-1}\underline{I}_{{I}}(q)\right)\\
&\,= \frac{\Pi_{W,0}(z)^M}{\alpha_1(z)^{M-1}\cdots\alpha_{M-1}(z)}\Delta_{M,N,I}(q)\,.
\end{align} 
It is easy to check that $\Delta_{M,N,I}(q) = 1$, and so find the explicit expression for the holomorphic solution:
\beq\label{eq:D_M_N}
D_{M}^{(W)}(z) := D_{(01\cdots(M-1))}^{(W)}(z) = \Pi_{W,0}(z)^M\,\prod_{j=1}^{M-1}\frac{1}{\alpha_{j}(z)^{M-j}}\,.
\eeq
This finishes the proof of eq.~\eqref{eq:det_to_Delta}. It is also easy to show that
\beq
D_{(01\cdots(M-2)M)}^{(W)}(z) = D_{(01\cdots(M-1))}^{(W)}(z)\,I(Y_{W,M-1};q)\,,
\eeq
in agreement with eq.~\eqref{eq:t_tilde_def}. Finally, we note that we can use eq.~\eqref{eq:D_M_N} to obtain an explicit expression for the structure series $\alpha_i(z)$ in terms of minors of the period matrix:
\beq
\alpha_i(z) = \frac{D_{i}^{({N+i-1})}(z)^2}{D_{i-1}^{({N+i-1})}(z)D_{i+1}^{({N+i-1})}(z)}\,,
\eeq
where we defined $D_{0}^{({N+i-1})}(z)=1$.